\documentclass[twocolumn,prX,showpacs,preprintnumbers,amsmath,amssymb,superscriptaddress,]{revtex4}

\usepackage{graphicx}
\usepackage{graphics}
\usepackage{dcolumn}
\usepackage{bm}
\usepackage{sidecap}
\usepackage{booktabs}

\begin{document}

\title{Mass measurements near the $r$-process path using the Canadian Penning Trap mass spectrometer}

\author{J. \surname{Van Schelt}}
 \affiliation{Department of Physics, University of Chicago, Chicago, Illinois 60637, USA}
 \affiliation{Physics Division, Argonne National Laboratory, Argonne, Illinois 60439, USA}
\author{D. Lascar}
 \affiliation{Department of Physics and Astronomy, Northwestern University, Evanston, Illinois 60208, USA}
 \affiliation{Physics Division, Argonne National Laboratory, Argonne, Illinois 60439, USA}
\author{G. Savard}
 \affiliation{Department of Physics, University of Chicago, Chicago, Illinois 60637, USA}
 \affiliation{Physics Division, Argonne National Laboratory, Argonne, Illinois 60439, USA}
\author{J. A. Clark}
 \affiliation{Physics Division, Argonne National Laboratory, Argonne, Illinois 60439, USA}
\author{S. Caldwell}
 \affiliation{Department of Physics, University of Chicago, Chicago, Illinois 60637, USA}
 \affiliation{Physics Division, Argonne National Laboratory, Argonne, Illinois 60439, USA}
\author{A. Chaudhuri}
 \affiliation{Department of Physics, University of Manitoba, Winnipeg, Manitoba R3T 2N2, Canada}
 \affiliation{Physics Division, Argonne National Laboratory, Argonne, Illinois 60439, USA}
 \altaffiliation[Now at]{TRIUMF, Vancouver, BC V6T 2A3, Canada}
\author{J. Fallis}
 \affiliation{Department of Physics, University of Manitoba, Winnipeg, Manitoba R3T 2N2, Canada}
 \affiliation{Physics Division, Argonne National Laboratory, Argonne, Illinois 60439, USA}
 \altaffiliation[Now at]{TRIUMF, Vancouver, BC V6T 2A3, Canada}
\author{J. P. Greene}
 \affiliation{Physics Division, Argonne National Laboratory, Argonne, Illinois 60439, USA}
\author{A. F. Levand}
 \affiliation{Physics Division, Argonne National Laboratory, Argonne, Illinois 60439, USA}
\author{G. Li}
 \affiliation{Department of Physics, McGill University, Montr\'{e}al, Qu\'{e}bec H3A 2T8, Canada}
 \affiliation{Physics Division, Argonne National Laboratory, Argonne, Illinois 60439, USA}
\author{K. S. Sharma}
 \affiliation{Department of Physics, University of Manitoba, Winnipeg, Manitoba R3T 2N2, Canada}
\author{M. G. Sternberg}
 \affiliation{Department of Physics, University of Chicago, Chicago, Illinois 60637, USA}
 \affiliation{Physics Division, Argonne National Laboratory, Argonne, Illinois 60439, USA}
\author{T. Sun}
 \affiliation{Physics Division, Argonne National Laboratory, Argonne, Illinois 60439, USA}
\author{B. J. Zabransky}
 \affiliation{Physics Division, Argonne National Laboratory, Argonne, Illinois 60439, USA}

\date{\today}

\begin{abstract}
The masses of 40 neutron-rich nuclides from Z~=~51 to 64 were measured at an average precision of $\delta m/m= 10^{-7}$ using the Canadian Penning Trap mass spectrometer at Argonne National Laboratory.  The measurements, of fission fragments from a $^{252}$Cf spontaneous fission source in a helium gas catcher, approach the predicted path of the astrophysical $r$ process.  Where overlap exists, this data set is largely consistent with previous measurements from Penning traps, storage rings, and reaction energetics, but large systematic deviations are apparent in $\beta$-endpoint measurements.  Differences in mass excess from the 2003 Atomic Mass Evaluation of up to $400$~keV are seen, as well as systematic disagreement with various mass models.
\end{abstract}

\pacs{21.10.Dr, 27.60.+j, 27.70.+q, 26.30.Hj}
\maketitle

\section{\label{sec:intro}Introduction}

Neutron-rich isotopes have become increasingly more accessible to experiments requiring accelerated beams or trapped ions in recent years, and regions long unexplored are now being probed.  Penning trap mass spectrometers are taking advantage of stopped fission and reaction products from gas catchers~\cite{Savard_IJMS, SHIPTRAP-146} and ISOL facilities~\cite{99Am05,ISOL-Cs08,Jyfl_Sr,Jyfl_Tc,ISOL-Xe09, TITAN-12Be} to give direct measurements of these nuclidic masses to high precision.  The masses of nuclides far from stability are of interest for a variety of fields, including astrophysics and nuclear structure~\cite{Bollen-review,Blaum-review}.

In particular, the rapid neutron-capture process ($r$ process)~\cite{B2FH,Cowan-rev,Arnould-rev} path is predicted to lie mostly in the region of unmeasured neutron-rich nuclides which is beginning to come into reach of precision study.  Determination of neutron separation energies ($S_n$) from nuclear masses is critical for establishing the path of the $r$ process, which is thought to lie near the line $S_n \approx 3$~MeV, and is important for numerical simulations of $r$-process dynamics in different environments~\cite{Arcones_simulation, Farouqi_simulation, Wanajo_simulation}.  Much of the path is still inaccessible to experiment, but extending mass measurements closer to it provides information to better constrain mass models and extrapolations to the proposed $r$-process path and further still to the neutron dripline.  Additionally, the appearance of neutron sub-shells, regions of deformation~\cite{deformation, Jyfl_Sr}, and shell quenching~\cite{quench} could be discovered via examination of $S_{2n}$ trends.

Continuing a previous program of fission fragment mass measurements~\cite{Savard_IJMS}, the Canadian Penning Trap (CPT) at Argonne National Laboratory has measured the masses of 40 neutron-rich nuclides from Z=51 to 64 near the target precision of $\frac{\delta m}{m} \approx 10^{-7}$, or \hbox{$\delta m\approx 15$~keV/$c^2$}.  This paper reports on the results and techniques of those measurements, and compares the results to previous measurements and the mass models on which $r$-process simulations depend.

\section{\label{sec:apparatus}Experimental Procedure}

\begin{figure}
\resizebox{0.49\textwidth}{!}{
\includegraphics{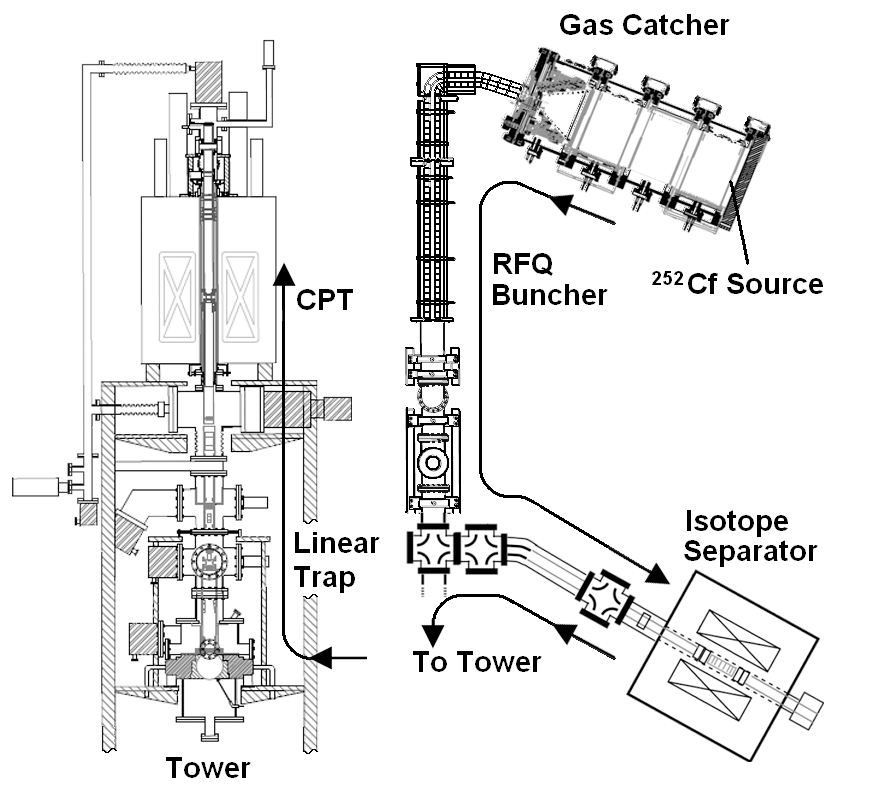}}
\caption{Layout of the portion of the CPT system used in these measurements.  Components are not to scale, and the CPT tower view has been separated from the rest of the system for convenience.  Arrows indicate direction of ion travel.}
\label{fig:system}
\end{figure}

The CPT and associated equipment have been described elsewhere~\cite{Savard_IJMS,Fallis_thesis}, but details emphasizing recent additions relevant to these measurements will be presented here.  The system, illustrated in Fig.~\ref{fig:system}, consists of a gas catcher for stopping reaction products, a radio frequency quadrupole (RFQ) cooler and buncher, an isotope separator Penning trap, and a linear Paul trap for accumulation and staging of ions before injection in the precision Penning trap.  Upgrades since the last CPT measurements in the neutron-rich region~\cite{Savard_IJMS} include a larger-volume gas catcher with a stronger fission source and a higher-resolution isotope separator in a superconducting magnet.

The gas catcher \cite{Savard_catcher} was designed to stop and cool ions produced either from an internal spontaneous fission source or from reactions using beams from the Argonne Tandem-Linac Accelerator System (ATLAS).  This work was performed with fission fragments from a \hbox{$150$-$\mu$Ci} $^{252}$Cf source placed behind a gold degrader foil with thickness optimized for collection of fragments in the heavy peak.  The $\approx 1$~m-length cylindrical gas catcher is filled with $50$\textendash$70$~torr of purified helium gas which stops the fission products through collision and ionization.  A large fraction, roughly $30$\textendash$50$\%, of these fission products stop as singly- or doubly-charged ions in the helium gas.  An electrostatic gradient placed along the catcher axis pushes the ions toward the exit nozzle, while an RF ion guide~\cite{Wada} keeps the ions from touching the walls and neutralizing.  At the downstream end the ion guide forms an RF funnel leading to the exit nozzle, where gas flow pushes the ions into the next stage.

Ions are carried away from the gas catcher nozzle by an RFQ ion guide~\cite{Herfurth}, and are separated from the helium gas by the combined effects of electric fields and differential pumping.  The ions are accumulated and bunched in a linear Paul trap at the end of the RFQ, where they are cooled further in $\approx 10^{-3}$~torr room-temperature helium gas.  For these measurements, after $100$~ms of accumulation the ion bunch is ejected from the trap into the beamline to the rest of the system.

Next, the ions are transferred to the isotope separator, where the ion bunches are purified.  The isotope separator is a gas-filled cylindrical Penning trap~\cite{Savard_gas} with a \hbox{$2.25$-T} superconducting magnet.  Penning traps allow two orbital motions of the trapped ions: the modified cyclotron motion with frequency $\omega_+$ and the much slower magnetron motion with frequency $\omega_-$, which are coupled by a cyclotron excitation $\omega_c=\omega_++\omega_-=qB/m$.  In the isotope separator, all ions are pushed outward by a mass-insensitive RF dipole field driven at $\omega_-$ while the ions of interest are re-centered in the trap by the application of a mass-sensitive RF quadrupole field at the $\omega_c$ frequency of those ions.  The result is the accumulation of a purified sample of the desired ions with a mass resolving power of approximately $5\times 10^3$ over the mass range studied here.  This is sufficient to remove ions of different mass number as well as any hydrocarbon contaminants that may be at the same mass number.  In cases where remaining molecular contamination is suspected, a strong dipole excitation at the reduced cyclotron frequency is applied to break up the molecules via gas collisions during the cleaning process.  After cleaning, ions are transferred to a second linear Paul trap where multiple bunches from the isotope separator are accumulated and cooled before transfer to the precision Penning trap.

The CPT mass spectrometer consists of a hyperbolic Penning trap~\cite{Geonium} in a highly stable \hbox{$5.9$-T} magnetic field and ultra-high vacuum, where masses are measured using the time-of-flight technique~\cite{Bollen_TOF}.  The main electrodes approximate hyperboloids of revolution, with apertures in the endcaps for ion entry and ejection and a splitting of the ring electrode into quadrants so that quadrupole excitations may be applied. Additional electrodes are placed between the endcaps and the ring to correct for the finite extent of the trap, and correction electrodes placed outside the endcaps prevent field penetration through the endcap apertures~\cite{Bollen_TOF}.

After an ion bunch is captured in this trap, the highest energy ions are evaporated away by briefly lowering the voltage of a correction tube and allowing them to escape.  This is done both to keep only those ions which are in the region of the trap with the most homogeneous magnetic field and to decrease the ion time-of-flight spread.  Any contaminants suspected to survive the isotope separator are then removed by a $100$- to $300$-ms dipole excitation at the $\omega_+$ of those ions.  Next, a dipole excitation at $\omega_-$ is applied for $40$~ms to position the ions in an orbit with the desired radius.

\begin{figure}
\resizebox{0.49\textwidth}{!}{
\includegraphics{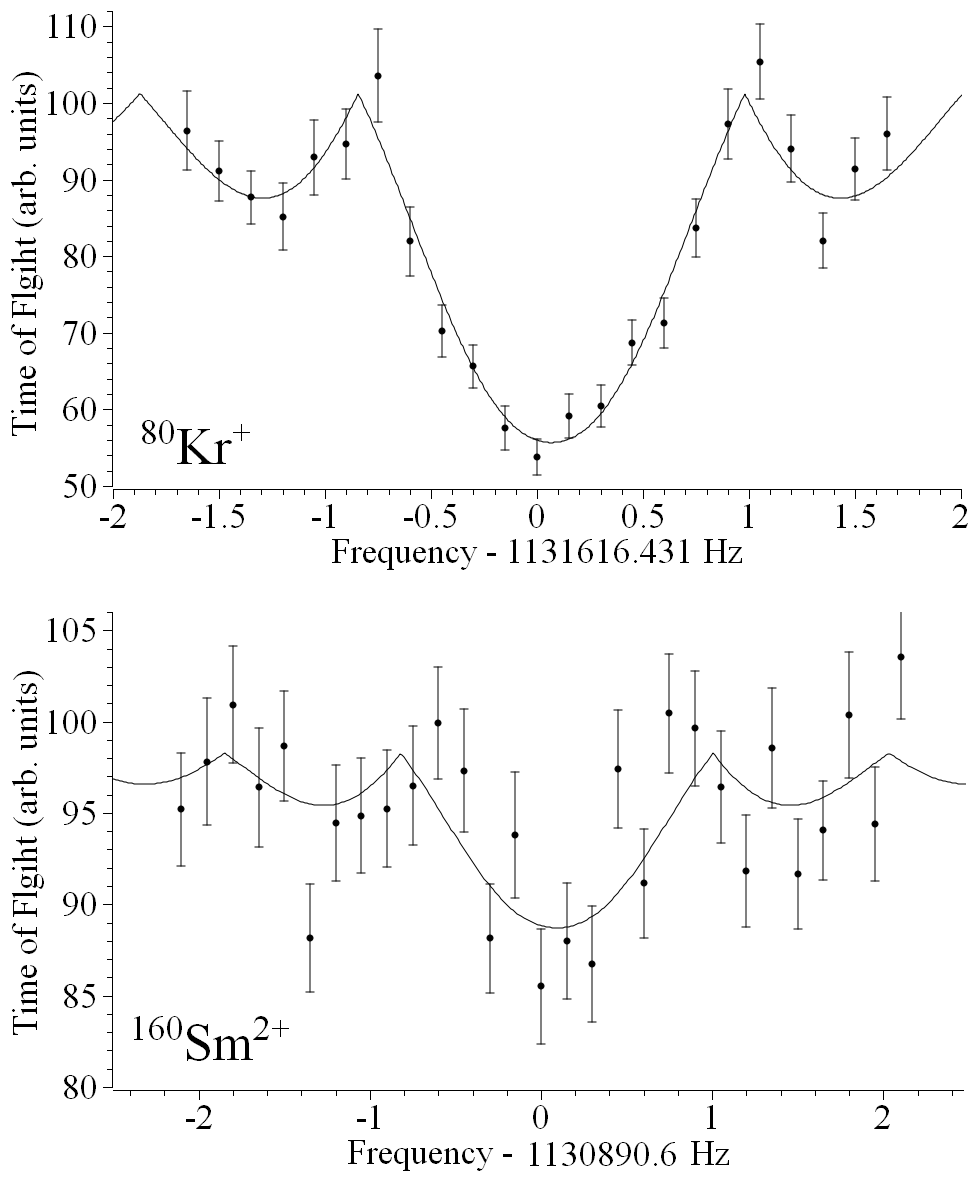}}
\caption{Two example cyclotron frequency scans from the CPT, for the calibrant $^{80}$Kr$^+$ and ion of interest $^{160}$Sm$^{2+}$.  Both excitations are 1~s long, and the fit function is a modified sinc function which approximates the true time of flight curve~\cite{Clark_thesis}.  The depth of the fit is less for the $^{160}$Sm$^{2+}$ scan because contaminant ions present in the Penning trap were not affected by the excitation.}
\label{fig:resonances}
\end{figure}

The last step inside the Penning trap is the application of a quadrupole excitation at a candidate cyclotron frequency for $200$ to $2000$~ms.  If the frequency applied matches the actual cyclotron frequency, the slow magnetron motion previously induced is converted to the much faster modified cyclotron motion at the same orbital radius~\cite{Konig}.  A typical $\frac{\omega_+}{\omega_-}$ ratio is $10^3$, thus the orbital kinetic energy increases by a factor of $10^6$.  After this final excitation the ion bunch is ejected from the trap and drifts down a \hbox{$\approx 1$-m} long beam line to either an MCP or channeltron time-of-flight detector outside the magnet.  As the ions travel through the gradient of the main magnetic field, the orbital motion is converted adiabatically to linear motion, accelerating the ions down the line.  Thus an application of a quadrupole field closer to the ion's true $\omega_c$ will result in a lower time of flight than a frequency farther away.  By scanning frequencies over successive bunches of ions, the minimum in time of flight can be found, and a measurement of the the ion's $\omega_c$ made.  Because the frequency is applied with a square amplitude envelope, the time-of-flight spectrum reflects that envelope's Fourier transform: a sinc function.  The cyclotron frequency of a calibrant ion of well-known mass is measured in the same manner, and the ratio of the frequencies is taken to cancel out the magnetic field term.  The mass of the neutral atom is then given by the calibrant mass and frequency ratio, with additional terms to compensate for the charge states and the masses of electrons not present given the charge states:
\[
m = \frac{\omega_{c(cal)}}{\omega_{c}} \frac{q}{q_{cal}} (m_{cal}-q_{cal}m_e) + qm_e
\]
where the subscript `cal' refers to the calibrant, $q$ is the integer charge state, and $m_e$ is the mass of the electron.  Alternatively, if multiple calibrant species are used, those frequency measurements can be combined as a single magnetic field strength measurement via $B=\omega_cm/q$, which can then be used to determine the unknown masses from their cyclotron frequency measurements. In either case, the calibrations need to be taken only as frequently as is required to monitor magnetic field drift, rather than after every measurement of an unknown ion.  Atomic binding energies are small enough to be neglected for these measurements.  Example time-of-flight scans are shown in Fig.~\ref{fig:resonances}.

\section{\label{sec:measurements}Measurements}

\begin{figure}
\resizebox{0.49\textwidth}{!}{
\includegraphics{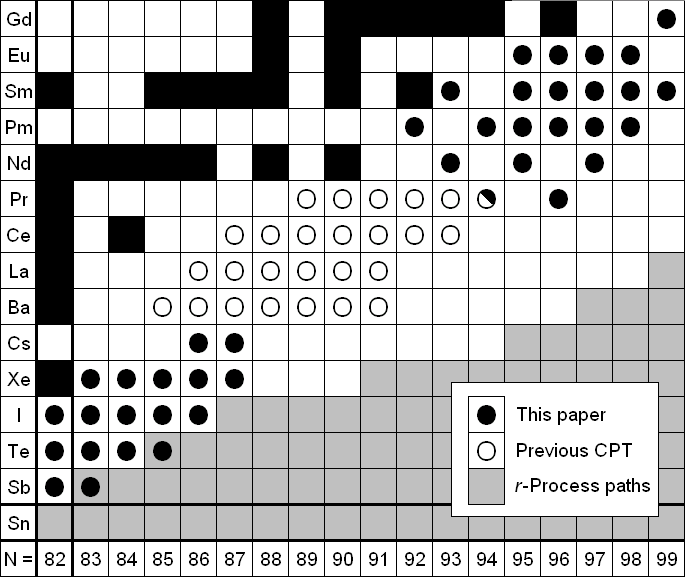}}
\caption{Positions of measured nuclides on the chart of the nuclides.  Nuclides marked with filled circles are those presented in this paper, and those with open circles were previously measured by the CPT mass spectrometer \cite{Savard_IJMS}.  The shaded area represents the span of several possible $r$-process paths~\cite{Arnould-rev}.}
\label{fig:chart}
\end{figure}

Following fission-fragment measurements made with a smaller gas catcher and source and a previous isotope separator~\cite{Savard_IJMS}, a new series of measurements in the $^{252}$Cf fission heavy peak began in April 2008.  The nuclides studied, identified in Fig.~\ref{fig:chart}, can be divided into two components:  the heaviest nuclides, from Pr to Gd, were measured in the $2^+$ charge state over $4~$weeks from April to July 2008; the lighter nuclides in the heavy peak, from Sb to Cs, were measured in the $1^+$ charge state over 4~weeks in February and March 2009.  High charge states are advantageous for mass measurements because their higher cyclotron frequencies lead to lower uncertainties in mass for the same uncertainty in frequency.  However, due to interactions with the helium gas used to stop and cool the ions in preparation for the precision Penning trap, they are limited to only $1^+$ or $2^+$ in practice.  The division of charge states in these measurements is due to the high ionization potential at the Xe electron shell closure, which allows the fission fragments with $Z \geq 56$ to survive as $2^+$, but not those with $Z <56$.

Excitation times used for candidate cyclotron frequencies range from 200 to 2000~ms, depending on the lifetime of the nuclide being studied and the conditions inside the trap.   The total duration of each measurement---typically 1 to 30~h---was dictated by the target precision of one part in $10^7$, the yield of ions out of the gas catcher, and the ultimate purity of the sample in the Penning trap.  To calibrate the $2^+$ measurements, which range in mass-to-charge ratio from $76.5$ to $81.5~\frac{\text{u}}{e}$, the cyclotron frequencies of $^{12}$C$_6^{~1}$H$_4^{~+}$,  $^{80}$Kr$^+$, and $^{86}$Kr$^+$ were measured approximately once per measurement week, and all of the calibration measurements were combined to give a single value for the magnetic field.  For the $1^+$ measurements, $^{136}$Xe$^+$ was measured at least once per measurement week.  In all weeks but one, at least one calibration resonance was taken with identical excitation time as used in that period to determine the fitting-function parameters to be used.  For $^{136}$I, no calibration was taken with the same excitation time as the measurement because of an unplanned interruption in the experiment.  To calibrate this \hbox{2000-ms} measurement, the results of a \hbox{1000-ms} calibration were used, and the width of the resonance was scaled from that calibration fit.  Table~\ref{tbl:calibrants} lists the mass values used for each calibrant, and Tables~\ref{tbl:1+_cal} and \ref{tbl:2+_cal} list the cyclotron frequency ratios of each measured ion to each calibration ion.

\begin{table}
\begin{tabular}{c r@{.}l c c}
\hline\noalign{\smallskip}
Calibrant & \multicolumn{2}{c}{Mass Used (u)} & Source & Relative weight\\
\noalign{\smallskip}\hline\noalign{\smallskip}
$^{136}$Xe$^+$   & 135&907\,214\,484(11)    & \cite{FSU_Xe} & 100\%\\
\noalign{\smallskip}\hline\noalign{\smallskip}
$^{12}$C$_6^{~1}$H$_4^{~+}$ & 76&031\,300\,085\,82(43) & \cite{AME03,Benzyne_heat,Benzyne_ion,Chem_ref} &  18\%\\
$^{80}$Kr$^+$     & 79&916\,3790(16)     & \cite{AME03} & 49\% \\
$^{86}$Kr$^+$     & 85&910\,610\,73(11)     & \cite{AME03} & 33\% \\
\end{tabular}
\caption{Mass values used for magnetic field calibrations.  For the molecule, the mass listed is that of the ion plus the mass of an electron.  The noble gas masses are those of neutral atoms.  Relative weights are given for the influence of each calibrant's mass on the corresponding set of measurements: $^{136}$Xe$^+$ for the $1^+$ and the others for $2^+$.}
\label{tbl:calibrants}
\end{table}

After the first week of $1^+$ measurements, an electrical discharge occurred in or near the precision Penning trap.  A lasting result of this discharge was an additional trap imperfection, possibly due to some surface charge deposited on a trap electrode.  Resonances taken under these conditions show an asymmetry, the effect of which had to be mitigated in the analysis as described below.  A study undertaken with various stable Xe isotopes~\cite{Scielzo_Te} found a systematic effect of only $0.37(33)~\frac{\text{keV}}{\text{u}}$ under these conditions, which is well below the statistical uncertainties in these measurements.

Most time-of-flight spectra were fit with a modified sinc function, which includes a parameter accounting for the possible over- or under-conversion from the $\omega_-$ to $\omega_+$ motions~\cite{Clark_thesis}.  For the data taken in the week following the discharge the time-of-flight spectrum showed an asymmetry in the side bands, pulling the modified sinc fit to the low-frequency side.  A Gaussian function was used to fit that week's data to minimize this problem, and the uncertainties were appropriately inflated due to the increased $\chi^2$ of these fits.  In all cases cuts were placed on the number of detected ions from a single ion bunch to minimize systematic effects discussed below.

\begin{figure}
\resizebox{0.49\textwidth}{!}{
\includegraphics{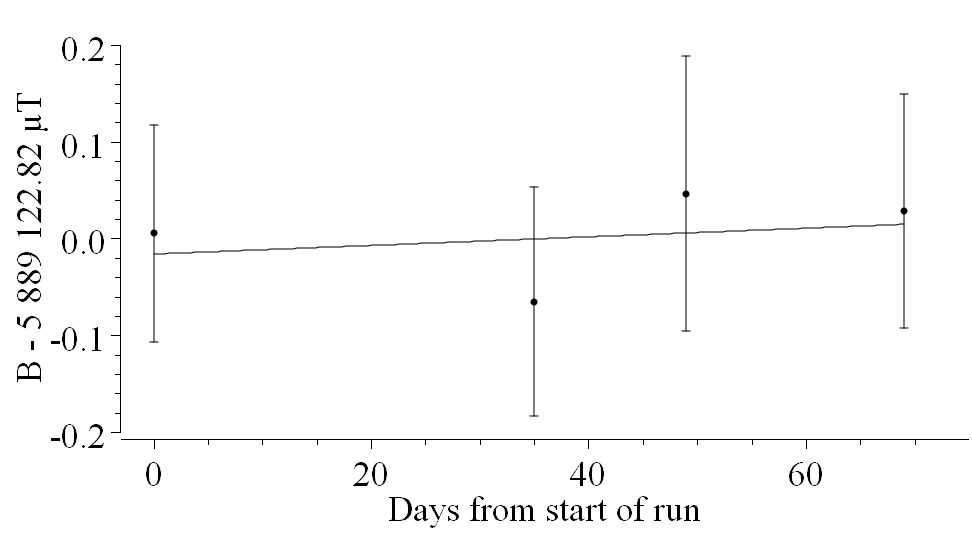}}
\caption{Results of magnetic field calibration data taken over the span of the $2^+$ measurements.  Each point represents the combined result over all calibration species taken during the respective measurement weeks.  The linear fit to these data gives a field drift of $0.1(0.4)$~ppb/day.  The consequence of this negligible drift is that no systematic correction nor uncertainty were necessary.  Note that not all weeks in this span contain mass measurements, but only those weeks for which calibrations are shown here.}
\label{fig:calibrants_time}
\end{figure}

There are several possible sources of systematic error, the largest of which are listed here.  As the magnetic field drifts, $\omega_c$ will change, but the \hbox{$5.9$-T} superconducting magnet is extremely stable (Fig.~\ref{fig:calibrants_time}), with the field drift measured to be $0.1(0.4)$~ppb/day over a two-month period.  Effects of trap misalignment and electric field imperfections are suppressed to high order~\cite{Gabby}.  In the weeks following the aforementioned discharge, the measured $\omega_c$ value for $^{136}$Xe$^+$ drifted up $0.3(0.6)$~ppb/day as the trap recovered.  This small potential source of error during this time was mitigated by using a calibration taken within one week of each measurement to calculate the mass.  The ion cloud's charge alters the electric field of the trap, but the effect on the $\omega_c$ value has been measured to be $< 2.5$~ppb per detected ion under normal conditions~\cite{Scielzo_Te}, and $< 17$~ppb per detected ion with the post-discharge effect during these measurements.  This effect will be suppressed if the trap population during measurement and calibration are kept low, as was the case here with the average number of detected ions per ejection ranging from 0.03 to 3.1 for the ions of interest, and below 7 for the calibration ions.  Contaminant ions of similar mass can have a separate effect as the ion motions interfere with each other.  Based on a 100 ppb limit at 15~ions~\cite{Bollen_isomer}, we estimate the effect to have an upper limit of $10$~ppb for the 3-ion maximum rate here.  The reference frequency used by the $\omega_c$ signal generator was compared to a Rb frequency standard and measured to be stable to $3$~ppb over a two-week period.  Mass-dependent effects in the precision Penning trap were investigated as part of a very high precision study of Xe isotopes~\cite{Scielzo_Te}.  A frequency-dependent attenuation of the applied $\omega_c$ amplitude was found at $-0.1\%$/kHz over the frequency range corresponding to $A/q$ from 113 to 165.  This, combined with a measured $\omega_c$ centroid dependence on amplitude due to the dodecapole moment of the trap, gives a variation of $-0.3~\frac{\text{keV}}{\text{u}}$, which is again mitigated by using calibrants of similar mass.  Because these potential sources of error are much smaller than our statistical uncertanties of at least 32~ppb, or  more typically 100~ppb, no systematic uncertainty was added to the quoted results.

\begin{table}
\begin{tabular}{l l c}
\noalign{\smallskip}\hline\hline\noalign{\smallskip}
\multicolumn{1}{c}{Ion}&\multicolumn{1}{c}{$\omega_c$ Ratio}&Excitation\\
&&Time (ms)\\
\noalign{\smallskip}\hline\noalign{\smallskip}
$^{133}$Sb$^+$&1.022\,510\,200(79)&1000\\
$^{134}$Sb$^+$&1.014\,832\,72(27)&500\\
\noalign{\smallskip}\hline\noalign{\smallskip}
$^{134}$Te$^+$&1.014\,904\,071(49)&1000\\
$^{135}$Te$^+$&1.007\,342\,824(75)&1000\\
$^{136}$Te$^+$&0.999\,905\,180(37)&1000\\
$^{137}$Te$^+$&0.992\,562\,30(13)&1000\\
\noalign{\smallskip}\hline\noalign{\smallskip}
$^{135}$I$^+$&1.007\,391\,356(62)&1000\\
$^{136}$I$^+$&0.999\,944\,000(36)&2000\\
$^{137}$I$^+$&0.992\,617\,351(65)&1000\\
$^{138}$I$^+$&0.985\,386\,600(46)&1000\\
$^{139}$I$^+$&0.978\,266\,90(22)&1000\\
\noalign{\smallskip}\hline\noalign{\smallskip}
$^{137}$Xe$^+$&0.992\,664\,183(83)&1000\\
$^{138}$Xe$^+$&0.985\,447\,843(42)&1000\\
$^{139}$Xe$^+$&0.978\,321\,228(81)&1000\\
$^{140}$Xe$^+$&0.971\,309\,239(75)&1000\\
$^{141}$Xe$^+$&0.964\,381\,577(69)&1000\\
\noalign{\smallskip}\hline\noalign{\smallskip}
$^{141}$Cs$^+$&0.964\,427\,61(15)&1000\\
$^{142}$Cs$^+$&0.957\,603\,370(75)&1000\\
\noalign{\smallskip}\hline\hline\noalign{\smallskip}
\end{tabular}
\caption{Ratios of the cyclotron frequencies of the ions of interest to those of the $^{136}$Xe$^+$ ions for measurements made in the $1^+$ charge state.  Calibrant mass used is listed in Table~\ref{tbl:calibrants}.}
\label{tbl:1+_cal}
\end{table}

\begin{table*}
\begin{tabular}{l l l l c}
\hline\hline\noalign{\smallskip}
&\multicolumn{3}{c}{$\omega_c$ Ratios}&Excitation\\
\noalign{\smallskip}\cline{2-4}\noalign{\smallskip}
\multicolumn{1}{c}{Ion}&\multicolumn{1}{c}{$^{76}(^{12}$C$_6^{~1}$H$_4)^+$}&\multicolumn{1}{c}{$^{80}$Kr$^+$}&\multicolumn{1}{c}{$^{86}$Kr$^+$}&Time (ms)\\
\noalign{\smallskip}\hline\noalign{\smallskip}
$^{153}$Pr$^{2+}$&0.994\,302\,50(26)&1.045\,110\,12(27)&1.123\,500\,53(29)&200\\
$^{155}$Pr$^{2+}$&0.981\,425\,59(20)&1.031\,575\,21(22)&1.108\,950\,41(23)&500\\
\noalign{\smallskip}\hline\noalign{\smallskip}
$^{153}$Nd$^{2+}$&0.994\,342\,931(34)&1.045\,152\,616(32)&1.123\,546\,210(39)&500\\
$^{155}$Nd$^{2+}$&0.981\,472\,30(11)&1.031\,624\,31(11)&1.109\,003\,19(12)&500\\
$^{157}$Nd$^{2+}$&0.968\,925\,46(29)&1.018\,436\,35(30)&1.094\,826\,04(32)&200\\
\noalign{\smallskip}\hline\noalign{\smallskip}
$^{153}$Pm$^{2+}$&0.994\,366\,01(15)&1.045\,176\,87(16)&1.123\,572\,28(17)&500\\
$^{155}$Pm$^{2+}$&0.981\,503\,965(56)&1.031\,657\,596(57)&1.109\,038\,971(63)&500\\
$^{156}$Pm$^{2+}$&0.975\,190\,689(43)&1.025\,021\,719(42)&1.101\,905\,358(48)&1000\\
$^{157}$Pm$^{2+}$&0.968\,964\,141(81)&1.018\,477\,002(84)&1.094\,869\,742(92)&500\\
$^{158}$Pm$^{2+}$&0.962\,807\,82(15)&1.012\,006\,10(16)&1.087\,913\,48(17)&500\\
$^{159}$Pm$^{2+}$&0.956\,733\,59(11)&1.005\,621\,48(12)&1.081\,049\,97(13)&500\\
\noalign{\smallskip}\hline\noalign{\smallskip}
$^{155}$Sm$^{2+}$&0.981\,526\,10(15)&1.031\,680\,86(16)&1.109\,063\,98(17)&500\\
$^{157}$Sm$^{2+}$&0.968\,993\,178(51)&1.018\,507\,523(52)&1.094\,902\,552(58)&500\\
$^{158}$Sm$^{2+}$&0.962\,848\,141(58)&1.012\,048\,482(58)&1.087\,959\,039(65)&1000\\
$^{159}$Sm$^{2+}$&0.956\,770\,122(65)&1.005\,659\,884(67)&1.081\,091\,252(74)&200 \& 500\\
$^{160}$Sm$^{2+}$&0.950\,775\,181(67)&0.999\,358\,608(69)&1.074\,317\,338(76)&1000\\
$^{161}$Sm$^{2+}$&0.944\,844\,874(74)&0.993\,125\,271(77)&1.067\,616\,458(84)&500\\
\noalign{\smallskip}\hline\noalign{\smallskip}
$^{158}$Eu$^{2+}$&0.962\,861\,30(15)&1.012\,062\,31(16)&1.087\,973\,91(17)&500\\
$^{159}$Eu$^{2+}$&0.956\,794\,898(63)&1.005\,685\,926(64)&1.081\,119\,248(71)&500\\
$^{160}$Eu$^{2+}$&0.950\,795\,89(10)&0.999\,380\,38(11)&1.074\,340\,74(12)&500\\
$^{161}$Eu$^{2+}$&0.944\,877\,14(11)&0.993\,159\,19(12)&1.067\,652\,92(13)&500\\
\noalign{\smallskip}\hline\noalign{\smallskip}
$^{163}$Gd$^{2+}$&0.933\,275\,822(91)&0.980\,965\,055(94)&1.054\,544\,14(10)&500\\
\noalign{\smallskip}\hline\hline

\end{tabular}
\caption{Ratios of the cyclotron frequencies of the ions of interest to those of the calibration ions for measurements made in the $2^+$ charge state.  Calibrant masses and relative weights used are listed in Table~\ref{tbl:calibrants}.  Uncertainties are those of calibrant and target ion combined.}
\label{tbl:2+_cal}
\end{table*}

\begin{table*}
\begin{tabular}{l@{\hspace{0.5cm}}l@{\hspace{0.5cm}}l@{\hspace{0.5cm}}l@{\hspace{0.5cm}}r@{}lll}
\hline\hline\noalign{\smallskip}
&\multicolumn{1}{c}{Mass (u)}&\multicolumn{6}{c}{Mass Excess (keV)}\\
\noalign{\smallskip}\cline{3-8}\noalign{\smallskip}
\multicolumn{1}{c}{Nuclide}&\multicolumn{1}{c}{CPT}&\multicolumn{1}{c}{CPT}&\multicolumn{1}{c}{AME03}&\multicolumn{2}{c}{$\Delta_{\textsc{CPT-AME03}}$}&\multicolumn{1}{c}{ISOLTRAP}&\multicolumn{1}{c}{FRS-ESR}\\
\noalign{\smallskip}\hline\noalign{\smallskip}
$^{133}$Sb&132.915\,277(10)&-78\,918.7(9.5)&-78\,943(25)&24&(27)&&-78\,986(120)\\
$^{134}$Sb$^{\ast}$&133.920\,812(35)&-73\,763(33)&-74\,170(40)&407&(52)&&\\
\noalign{\smallskip}\hline\noalign{\smallskip}
$^{134}$Te&133.911\,3976(65)&-82\,532.6(6.0)&-82\,559(11)&26&(13)&&-82\,758(121)\\
$^{135}$Te&134.916\,550(10)&-77\,733.2(9.3)&-77\,830(90)&97&(90)&&-77\,725(123)\\
$^{136}$Te&135.920\,1024(50)&-74\,424.2(4.6)&-74\,430(50)&6&(50)&&\\
$^{137}$Te&136.925\,622(18)&-69\,282(17)&-69\,560(120)&280&(120)&&-69\,290(120)\\
\noalign{\smallskip}\hline\noalign{\smallskip}
$^{135}$I&134.910\,0503(82)&-83\,787.6(7.7)&-83\,790(7)&2&(10)&&\\
$^{136}$I$^{\ast}$&135.914\,8257(49)&-79\,339.3(4.5)&-79\,500(50)&161&(50)&&\\
$^{137}$I&136.918\,0282(90)&-76\,356.2(8.3)&-76\,503(28)&147&(29)&&-76\,518(121)\\
$^{138}$I&137.922\,7265(64)&-71\,979.8(6.0)&-72\,330(80)&350&(80)&&\\
$^{139}$I&138.926\,506(31)&-68\,460(29)&-68\,840(30)&380&(42)&&-68\,527(121)\\
\noalign{\smallskip}\hline\noalign{\smallskip}
$^{137}$Xe&136.911\,569(11)&-82\,373(11)&-82\,379(7)&6&(13)&-82\,382.2(1.8)&\\
$^{138}$Xe&137.914\,1550(59)&-79\,964.1(5.5)&-80\,150(40)&186&(40)&-79\,975.1(3.3)&\\
$^{139}$Xe&138.918\,791(11)&-75\,645(11)&-75\,644(21)&-1&(24)&-75\,644.6(2.1)&\\
$^{140}$Xe&139.921\,658(11)&-72\,976(10)&-72\,990(60)&14&(61)&-72\,986.5(2.3)&-72\,870(121)\\
$^{141}$Xe&140.926\,785(10)&-68\,199.5(9.4)&-68\,330(90)&130&(90)&-68\,197.3(2.9)&-68\,521(127)\\
\noalign{\smallskip}\hline\noalign{\smallskip}
$^{141}$Cs&140.920\,058(21)&-74\,466(19)&-74\,477(11)&11&(22)&-74\,475(15)$^\ddag$&\\
$^{142}$Cs&141.924\,303(11)&-70\,511(10)&-70\,515(11)&4&(15)&-70\,521(15)$^\ddag$&\\
\noalign{\smallskip}\hline\hline \noalign{\smallskip}
$^{153}$Pr$^\dag$&152.933\,895(15)&-61\,576(14)&-61\,630(100)&54&(100)&&\\
$^{155}$Pr&154.940\,508(32)&-55\,416(30)&-55\,780(300)$\#$&360&(300)$\#$&&\\
\noalign{\smallskip}\hline\noalign{\smallskip}
$^{153}$Nd&152.927\,7156(47)&-67\,332.5(4.4)&-67\,349(27)&16&(27)&&\\
$^{155}$Nd&154.933\,134(17)&-62\,285(16)&-62\,470(150)$\#$&190&(150)$\#$&&\\
$^{157}$Nd&156.939\,383(46)&-56\,464(43)&-56\,790(200)$\#$&330&(200)$\#$&&\\
\noalign{\smallskip}\hline\noalign{\smallskip}
$^{153}$Pm&152.924\,167(24)&-70\,638(22)&-70\,685(11)&47&(25)&&\\
$^{155}$Pm&154.928\,1350(85)&-66\,941.8(7.9)&-66\,970(30)&28&(31)&&\\
$^{156}$Pm&155.931\,1155(64)&-64\,165.5(5.9)&-64\,220(30)&54&(31)&&\\
$^{157}$Pm&156.933\,119(13)&-62\,299(12)&-62\,370(110)&70&(110)&&\\
$^{158}$Pm&157.936\,563(24)&-59\,091(23)&-59\,090(130)&0&(130)&&\\
$^{159}$Pm&158.939\,284(19)&-56\,557(17)&-56\,850(200)$\#$&290&(200)$\#$&&\\
\noalign{\smallskip}\hline\noalign{\smallskip}
$^{155}$Sm&154.924\,642(24)&-70\,196(22)&-70\,197.2(2.6)&1&(22)&&\\
$^{157}$Sm&156.928\,4166(80)&-66\,679.5(7.4)&-66\,730(50)&50&(51)&&\\
$^{158}$Sm&157.929\,9497(91)&-65\,251.5(8.5)&-65\,210(80)&-42&(80)&&\\
$^{159}$Sm&158.933\,215(10)&-62\,209.6(9.7)&-62\,210(100)&0&(100)&&\\
$^{160}$Sm&159.935\,333(11)&-60\,237(10)&-60\,420(200)$\#$&180&(200)$\#$&&\\
$^{161}$Sm&160.939\,158(12)&-56\,674(12)&-56\,980(300)$\#$&310&(300)$\#$&&\\
\noalign{\smallskip}\hline\noalign{\smallskip}
$^{158}$Eu&157.927\,791(25)&-67\,262(23)&-67\,210(80)&-52&(83)&&\\
$^{159}$Eu&158.929\,100(10)&-66\,043.2(9.5)&-66\,053(7)&10&(12)&&\\
$^{160}$Eu&159.931\,849(17)&-63\,482(16)&-63\,370(200)$\#$&-110&(200)$\#$&&\\
$^{161}$Eu&160.933\,662(19)&-61\,793(18)&-61\,780(300)$\#$&0&(300)$\#$&&\\
\noalign{\smallskip}\hline\noalign{\smallskip}
$^{163}$Gd&162.934\,175(16)&-61\,316(15)&-61\,490(300)$\#$&170&(300)$\#$&&\\
\noalign{\smallskip}\hline\hline\noalign{\smallskip}
\multicolumn{8}{l}{ $^{\ast}$: Identity of measured state is ambiguous due to the possible presence of an isomer.}\\
\multicolumn{8}{l}{ $\dag$: Combined CPT result with value from \cite{Savard_IJMS}.  See Sec.~\ref{sec:nuclides} for details.}\\
\multicolumn{8}{l}{ $\ddag$: Result has been adjusted due to a change in the calibration value since the publication of~\cite{99Am05}.}\\
&\multicolumn{7}{l}{See Sec.~\ref{sec:nuclides} for details.}\\
\multicolumn{8}{l}{ $\#$: Extrapolated mass value.}\\
\end{tabular}

\caption{Mass measurement results from the CPT.  Results are given as the masses of neutral atoms.  Sb through Cs were measured in the $1^+$ charge state, while Pr through Gd were measured in $2^+$.  Also listed are the mass excesses from the AME03~\cite{AME03}, the differences between the CPT and AME03 values, and the measurements from ISOLTRAP~\cite{99Am05,ISOL-Xe09} and FRS-ESR~\cite{GSI_08}.}
\label{tbl:masses}
\end{table*}

\section{\label{sec:discussion}Discussion}

Of the 40 nuclides measured, 8 had been measured previously by Penning traps, 20 by $\beta$-endpoint only, and 3 had no previous mass measurement of any kind.  Measurement uncertainties in the CPT results range from 5 to 46~${\normalfont \mu}$u, with most below or near the target $10^{-7}$ fractional uncertainty.  Isomers with lifetimes long enough to be captured in the trap are not expected for any of the measured nuclides except $^{134}$Sb~\cite{02Ko53, Shergur} and $^{136}$I~\cite{80KeZQ}, for which the identities of the measured states are unclear.

\begin{figure}
\resizebox{0.49\textwidth}{!}{
\includegraphics{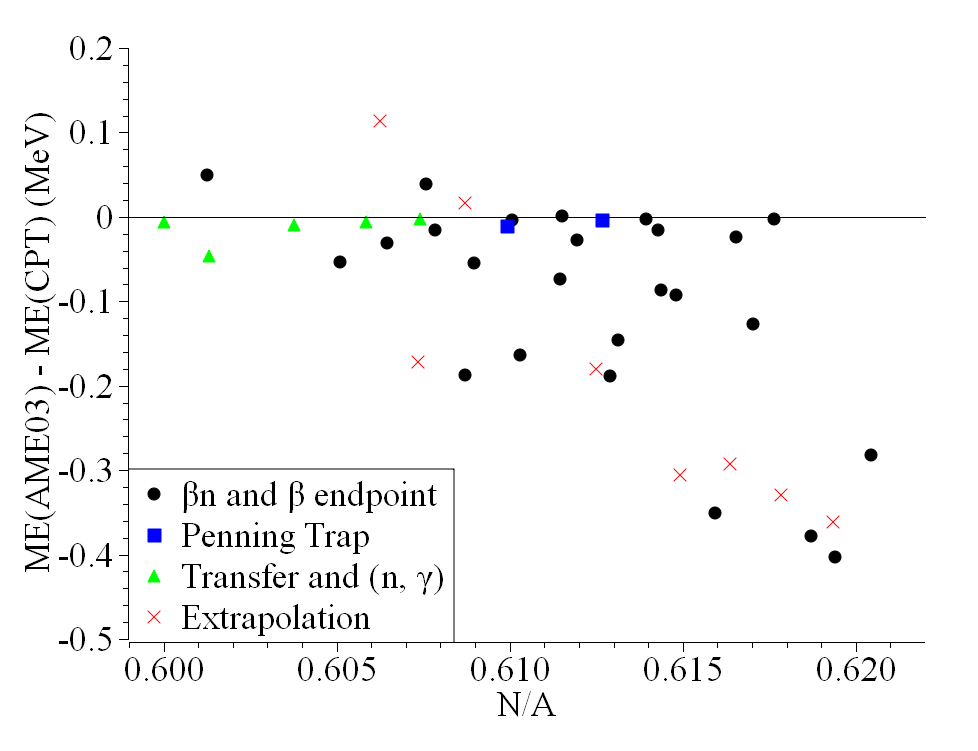}}
\caption{(Color online) The difference between mass excess values from the AME03~\cite{AME03} and CPT results.  The horizontal axis was chosen to illustrate the increase in deviation with distance from stability.  For each nuclide, the measurement technique most heavily weighted in the AME03 is indicated by the symbol used for each point.  In cases where data from multiple techniques were used the most heavily weighted technique is indicated.  Error bars are omitted for clarity; $\beta$-endpoint uncertainties range from 11 to 153~keV, Penning trap uncertainties are 11~keV, transfer and capture uncertainties range from 2.6 to 11~keV, and extrapolated mass uncertainties range from 150 to 300~keV.}
\label{fig:AME-ME}
\end{figure}

Table \ref{tbl:masses} shows the new CPT measurements and compares them to the 2003 Atomic Mass Evaluation (AME03)~\cite{AME03} as well as some more recent measurements~\cite{99Am05,ISOL-Xe09,GSI_08}.  The overall trend of these measurements versus the AME03 is of increasing mass as distance from stability increases, as seen in Fig.~\ref{fig:AME-ME}.  This effect was also seen in the CPT's previous measurements of Ba, La, Ce, and Pr fission fragments~\cite{Savard_IJMS}, proton-rich nuclides of Nb to Rh~\cite{Fallis_90}, as well as measurements by other Penning traps~\cite{Jyfl_Sr,Jyfl_Tc,ISOL-Xe09}.  Because the source data for the AME and its extrapolations are mostly from $\beta$-endpoint measurements, the observed deviations suggest an unaddressed systematic problem with that technique.  The frequency of such disagreement suggests that $\beta$-endpoint results as a whole should not be trusted as a reliable source of data for extrapolations.

\subsection{Comparison with past measurements}

The trends evident in Fig.~\ref{fig:AME-ME} demand separate comparisons between the CPT and literature mass values for different measurement techniques.  Four categories have been selected for individual discussion below, grouped so that systematic trends in differences may be identified and each method evaluated for accuracy.  Following these is a brief comparison of each new CPT measurement with the literature values for that nuclide.

\subsubsection{Penning Traps}

\begin{figure}
\resizebox{0.49\textwidth}{!}{
\includegraphics{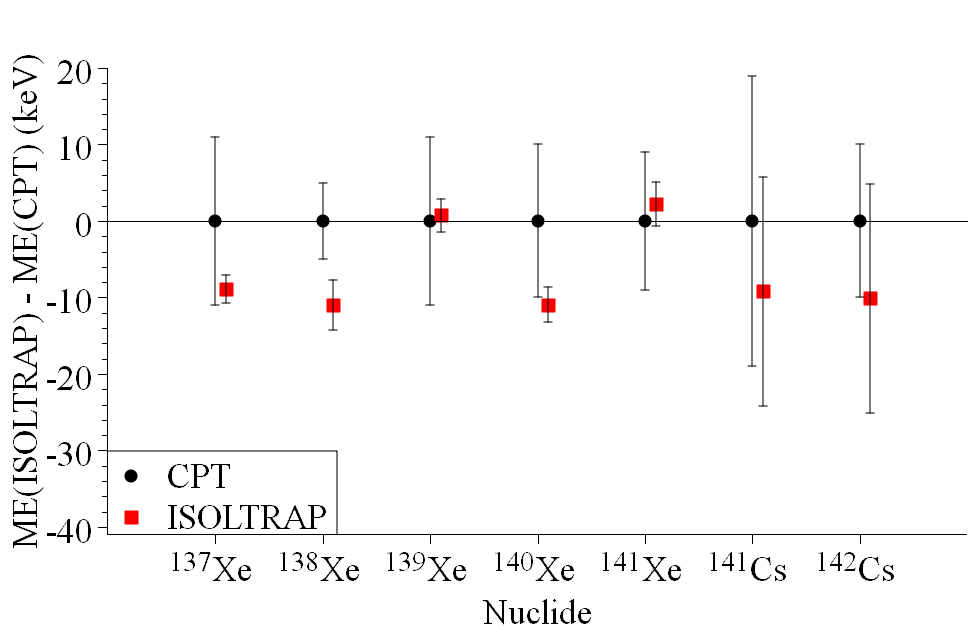}}
\caption{(Color online) Comparison of CPT and ISOLTRAP mass excess values for Xe~\cite{ISOL-Xe09} and Cs~\cite{99Am05} isotopes.  No significant individual or systematic differences are seen.}
\label{fig:ISOL-CPT}
\end{figure}

Some of the masses measured here have been previously measured in the ISOLTRAP or CPT Penning traps.  These are ideal checks on the accuracy of the experiment as a whole.  The CPT had previously measured and published the mass of $^{153}$Pr~\cite{Savard_IJMS}, and the present measurement is in agreement. ISOLTRAP has measured isotopic chains of Cs~\cite{99Am05} and Xe~\cite{ISOL-Xe09} from proton-induced fission at ISOLDE to neutron numbers beyond the current reach of the CPT for these elements.  Fig.~\ref{fig:ISOL-CPT} shows a comparison between the seven nuclides measured by both traps, and there is no evidence of any systematic or isolated differences, with $\chi^2/6 = 0.9$.  On average, the reported ISOLTRAP masses are $0.61(67)~\sigma$ lower than the CPT masses.

\subsubsection{Storage Rings}

\begin{figure}
\resizebox{0.49\textwidth}{!}{
\includegraphics{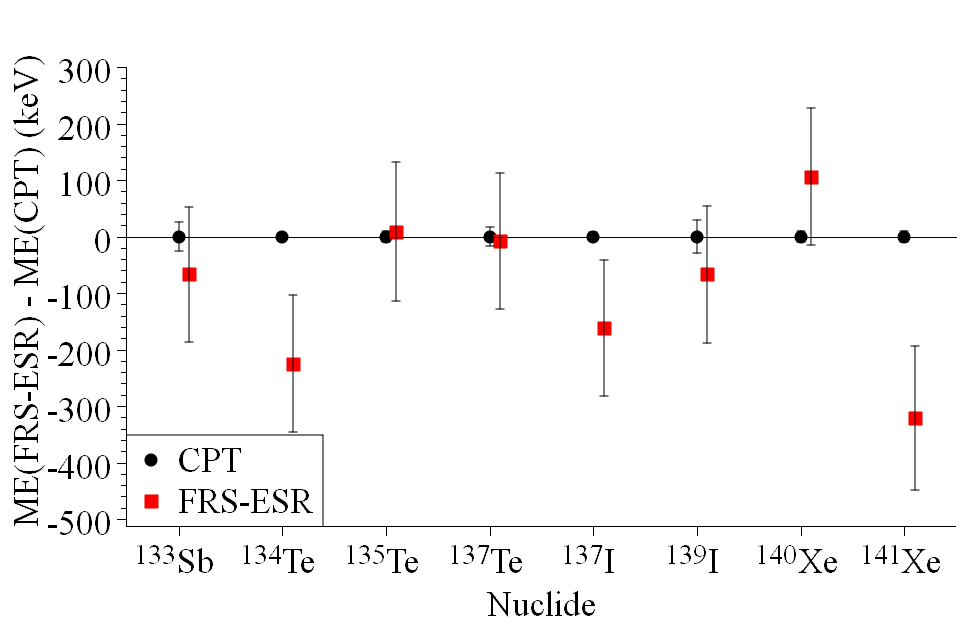}}
\caption{(Color online) Comparison of CPT and FRS-ESR mass excess values~\cite{GSI_08}.  In some cases the CPT error bars are smaller than the points.}
\label{fig:ESR-CPT}
\end{figure}

The FRS-ESR~\cite{GSI_08} cooler-storage ring at GSI uses isochronous mass spectrometry to measure masses of fission products in a radioactive beam via time of flight.  Eight nuclides of Sb, Te, I, and Xe have been measured at both the FRS-ESR and the CPT, and are compared in Fig.~\ref{fig:ESR-CPT}.  The two methods are in rough agreement but there is some scatter with an RMS difference of $158$~keV, despite the \hbox{$120$-keV} uncertainty typical of the FRS-ESR results.  Comparing the two data sets yields $\chi^2/7=1.8$, giving a statistical $p$-value of $3\%$ .  

If this large scatter is a real effect, then---given the strong consistency of the Penning traps discussed above---it may be due to a flaw in either the FRS-ESR experiment or calibrant mass values.  This last possibility is a significant one, given that some of those calibrant masses are largely determined via $\beta$-endpoint.  Unfortunately, not enough information is given in the FRS-ESR publication~\cite{GSI_08} to accurately recalculate new mass values with updated calibrant masses and compare to the CPT results.

\subsubsection{Transfer and Capture Reactions}

Five of the masses presented here had previously been established from nucleon-transfer or neutron-capture experiments.  The $(n,\gamma)$ reaction offers an opportunity for exquisitely precise neutron-separation energy ($S_n$) measurements due to the monoenergetic gamma rays and the state of gamma-ray absorption detector technology.  $S_n$($^{137}$Xe) has been measured to a precision of $80$~eV by this method~\cite{Tecdoc}, in agreement with the CPT result. $S_n$($^{155}$Sm) has twice been measured to sub-keV precision~\cite{82Ba15,82Sc03} and these are consistent, both with each other and with the CPT result.

Transfer reactions offer similar benefits, and four such measurements have been made on nuclides presented here.  The mass of $^{135}$I has been measured via $^{136}$Xe$(d,^3$He$)$~\cite{71Wi04}, $^{153}$Pm via $^{154}$Sm$(d,^3$He$)$~\cite{76Sh.B} and $^{154}$Sm$(t,\alpha)$~\cite{78Bu18}, and $^{159}$Eu via $^{160}$Gd$(t,\alpha)$~\cite{79Bu05}, to precisions of $40$, $25$, $20$, and $8$~keV/$c^2$, respectively.  The differences between these and the CPT measurements, using AME03 mass values for the Sm and Gd parent nuclides, are $0.02$, $0.12$, $1.5$, and $1.2~\sigma$, respectively, confirming the accuracy and reliability of transfer reaction measurements.

\begin{figure}
\resizebox{0.49\textwidth}{!}{
\includegraphics{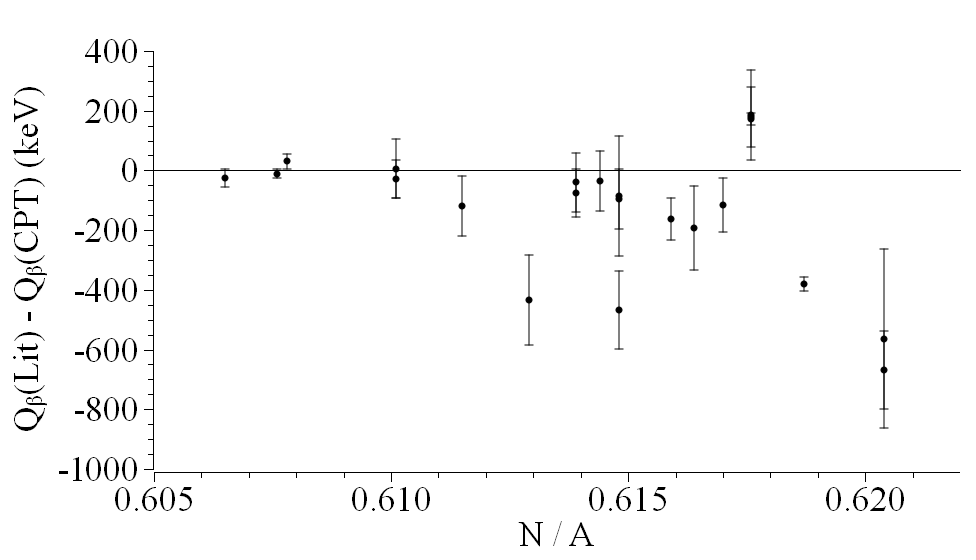}}
\caption{Comparison of $Q_\beta$ values as determined by $\beta$-endpoint measurements and CPT mass measurements of parent and daughter masses.  The apparent trend is to more negative difference in the $\beta$-endpoint measurements with distance from stability.  The outlier from this trend at $N/A=0.618$ is $^{136}$Te.}
\label{fig:Q-betas}
\end{figure}

\subsubsection{$\beta$-Endpoint Measurements}

$\beta$-endpoint measurements dominate the AME03 on the neutron-rich side of stability because of the applicability of that kind of measurement with a small number of nuclei.  For 15 of the nuclides presented here where isomeric states are not suspected, the $\beta$-decay $Q$-values can be calculated entirely from these CPT measurements, and a total of 22 $\beta$-endpoint measurements exist of these nuclides. A comparison of these data sets shows a systematic problem with the $\beta$-endpoint method, as illustrated in Fig.~\ref{fig:Q-betas}.  Of the 22 measurements, 10 are within $1~\sigma$ and another 6 within $2~\sigma$, but a long tail exists up to $9~\sigma$.  Comparison of the data sets gives $\chi^2/21 = 10.7$, which has a statistical $p$-value of $6\times 10^{-36}$.  Of the 22 $\beta$-endpoint measurements, 17 had values of  $Q_\beta$ smaller than the CPT results, which gives support to the notion that feeding to higher-lying states is erroneously pushing these measurements to lower masses.  

Two recent sets of measurements demonstrate the importance of considering systematic uncertainties in $\beta$-endpoint experiments.  Hayashi \emph{et al.}~\cite{Hayashi} give new $Q_\beta$ measurements for some of the Eu and Gd nuclides presented here and a new analysis of that group's earlier Pm and Sm~\cite{02Sh.A, Shibata03} measurements, and Fogelberg \emph{et al.}~\cite{07Fo02} measured $Q_\beta$ for neutron-rich Te and I isotopes.  The CPT results have good statistical agreement with the Hayashi \emph{et al.} results, but less so with Fogelberg \emph{et al.}, where deviations up to $9~\sigma$ are seen.  However, Fogelberg \emph{et al.} report only a statistical uncertainty, and if the \hbox{$100$-keV} systematic uncertainty suggested by Shibata \emph{et al}.~\cite{Shibata_unc} and used by Hayashi \emph{et al.} for the root-plot analysis method is applied, the disagreements decrease to less than $2~\sigma$.

\subsubsection{Discussion by Nuclide}\label{sec:nuclides}

What follows is a comparison to previous measurements of each nuclide presented here, ordered by element:

\textbf{$^{133}$Sb:} The AME03 cites a single experiment for the mass of $^{133}$Sb, a $\beta$-endpoint measurement~\cite{99Fo01}.  The value of $Q_\beta$ in that paper is $4.002(7)$~MeV, leading to an evaluated mass excess of $-78.943(25)$~MeV.  This is consistent with the CPT value of $-78.9187(95)$~MeV.  A more recent experiment by the FRS-ESR facility measured $^{133}$Sb via time of flight, and reported a mass excess of $-78.986(120)$~MeV~\cite{GSI_08}, which is also consistent with the CPT value.

\textbf{$^{134}$Sb:} There is a metastable state of $^{134}$Sb with a \hbox{10-s} lifetime and excitation energy estimated at $250$~keV~\cite{02Ko53}, and with a measurement from $\gamma$ coincidence at 279~keV~\cite{Shergur}.  Only one resonance was observed with the CPT in the expected mass region, so the state of the measured ions is unclear.  If the two states are within 400~keV of each other they would not be separately resolved in the CPT given the excitation time of 500~ms, so the observed resonance may be some weighted average of the two states.  

The AME03 cites two $\beta$-endpoint measurements for the mass determination of $^{134}$Sb: $Q_\beta = 8.390(45)$~MeV~\cite{95Me16} and $8.420(120)$~MeV~\cite{87Gr.A}, both of which are significantly lower (6.9 and 2.8 $\sigma$) than the CPT value of $8.769(34)$~MeV.  If the measured state is  assumed to be the metastable state, then a combination of the CPT measurement and the $\beta$-endpoint measurements above gives an excitation energy of $379(56)$~keV.  We do not offer an assignment of the measured state.

\textbf{$^{134}$Te:} A single $\beta$-endpoint measurement is cited by the AME03 with $Q_\beta = 1.513(7)$~MeV~\cite{99Fo01}, leading to a mass excess of $-82.559(11)$~MeV.  This is marginally consistent with our result of $-82.5326(60)$~MeV.  The FRS-ESR measurement gives $-82.758(121)$~MeV~\cite{GSI_08}, a $1.9~\sigma$ difference from the CPT.

\textbf{$^{135}$Te:} Two $\beta$-endpoint measurements were used in the AME03 to calculate the mass excess of $^{135}$Te: $Q_\beta = 5.960(100)$~MeV~\cite{87Gr.A}, and $5.970(200)$~MeV~\cite{85Sa15}.  These are combined in the AME03 to give $5.960(90)$~MeV, consistent with the value from CPT measurements of $6.054(12)$~MeV.  The recent FRS-ESR result gives a mass excess of $-77.725(123)$~MeV~\cite{GSI_08}, which is consistent with the CPT value of $-77.7332(93)$~MeV.

A $\beta$-endpoint result by Fogelberg \emph{et al.}~\cite{07Fo02} more recent than the AME03 gives $Q_\beta = 5.888(13)$~MeV, a 9.4~$\sigma$ difference from the CPT.  It is important to note that there was no systematic uncertainty assigned to that $Q_\beta$ measurement, only a statistical one.  Other authors~\cite{Shibata_unc} suggest a \hbox{100-keV} systematic uncertainty for the root-plot method employed in that study.  Increasing the Fogelberg \emph{et al.} uncertainty to 100~keV reduces the disagreement to $1.6~\sigma$.

\textbf{$^{136}$Te:} Three measurements are used by the AME03 to determine this mass.  The determination is dominated by a beta-delayed neutron-emission $Q$-value measurement of $Q_{\beta n} = 1.285(50)$~MeV~\cite{84Kr.B}, which agrees with the CPT value of $1.292(9)$~MeV.  The other citations are of beta-decay $Q$-value measurements of $Q_{\beta} = 5.095(100)$~MeV~\cite{87Gr.A} and $5.100(150)$~MeV~\cite{77Sc21}.  Assuming the CPT $^{136}$I measurement is of the ground state (see discussion below) both of these are somewhat higher than the CPT result of $4.915(7)$~MeV.

Since the AME03 publication, a new $\beta$-endpoint measurement has been made with $Q_{\beta} = 5.086(20)$~MeV~\cite{07Fo02}, a disagreement of 171~keV and $8.1~\sigma$ with the CPT result.  Adding a \hbox{$100$-keV} systematic uncertainty to the $\beta$-endpoint method as discussed above decreases the inconsistency to $1.7~\sigma$.

\textbf{$^{137}$Te:} Two $\beta$-endpoint experiments are cited in the AME03 for $^{137}$Te: $Q_\beta = 7.030(300)$~MeV~\cite{85Sa15} and $6.925(130)$~MeV~\cite{87Gr.A}.  These are combined in the AME03 to $6.940(120)$~MeV, marginally consistent with the CPT result of  $7.074(19)$.  The AME03 mass excess value of $-69.560(120)$~MeV is 278~keV lower than our value of $-69.282(17)$~MeV.  Over half of this difference, 147~keV, comes from the difference between the mass of $^{137}$I used by the AME03 and that determined in our experiment, discussed below.  The recent FRS-ESR measurement agrees with the CPT, with mass excess $-69.290(120)$~MeV~\cite{GSI_08}.

\textbf{$^{135}$I:} In the AME03, two different methods primarily constrain the $^{135}$I mass:  a $\beta$-endpoint measurement of $Q_\beta=2.627(6)$~MeV~\cite{99Fo01} and a $Q$($^{136}$Xe($d$,$^3$He)$^{135}$I) measurement of $-4.438(40)$~MeV~\cite{71Wi04}.  The CPT $^{135}$I measurement gives $Q_{(d,^3\textrm{He})}(^{136}$Xe$)=-4.437(8)$~MeV, in excellent agreement with the AME03 input value.  There is also a slight contribution to the $^{135}$I mass in the AME03 from a $^{136}$Te $Q_{\beta n}$ measurement~\cite{84Kr.B} of $1.285(50)$~MeV, consistent with CPT's $1.292(9)$~MeV.  The CPT results cannot be directly compared to the $Q_\beta$ measurement.

\textbf{$^{136}$I:} There is a metastable state of $^{136}$I~\cite{80KeZQ} which, with the ground state, is expected to be produced in fission.  Only one  resonance was clearly measured during this experiment, and its identity is uncertain.  If the states lie within 100~keV of each other they would not have been resolved given the excitation time of 2000~ms.  The determination of the number of states trapped by the CPT was made difficult by the possible presence of $^{135}$Xe$^1$H$^+$ less than 1~Hz (200~keV) away from the observed $^{136}$I or $^{136}$I$^m$ resonance.  Because the mass of $^{135}$Xe$^1$H$^+$ is well known we can be certain that the resonance measured is not that molecule.  The mass excess of the observed $^{136}$I state is found to be $-79.3393(45)$~MeV, and $Q_\beta$ is $7.0898(45)$~MeV.

The AME03 cites three papers for a total of four measurements of $^{136}$I and its metastable state, each a $\beta$-endpoint measurement.  The most recent of those~\cite{87Gr.A} gives $Q_\beta = 6.925(70)$~MeV and $7.705(120)$~MeV for the two states.  An older paper~\cite{76Lu04} was disregarded for its measurement of the ground-state $Q_\beta$~value, but its measurement of $7.100(230)$~MeV assigned to the isomer state was included in the evaluation.  A third paper~\cite{59Jo37} gives only the ground state $Q_\beta$~value at $6.960(100)$~MeV.  The adopted AME03 $Q_\beta$~values are $6.930(50)$~MeV for the ground state and $7.580(110)$~MeV for the metastable state.

These suggest that it is the ground state which was measured here, however, a paper published since the AME03 claims $\beta$-endpoint measurements of both the ground and metastable states, with $Q_\beta = 6.850(20)$ and $7.051(12)$~MeV~\cite{07Fo02}, respectively.  This is the same paper discussed above which did not include systematic uncertainties.  Given the conflicting measurements it is difficult to assign a state to the measured ion.  

\textbf{$^{137}$I:} The AME03 cites a $\beta$-delayed neutron study for this mass determination, with $Q_{\beta n}=1.850(30)$~MeV~\cite{84Kr.B}.  The CPT measurement gives $Q_{\beta n} = 2.001(8)$~MeV, which disagrees by $5~\sigma$.  Disregarded by the AME03 is a $\beta$-endpoint measurement of $Q_\beta=5.880(60)$~MeV~\cite{87Gr.A}, which is inconsistent with the CPT result of $6.017(14)$~MeV.  The recent FRS-ESR measurement gave a mass excess of $-76.518(121)$~MeV~\cite{GSI_08}, which differs from the CPT result of $-76.3562(83)$~MeV by $1.3~\sigma$.

\textbf{$^{138}$I:} Only one experiment is cited in the literature for $^{138}$I, a $\beta$-endpoint measurement of $Q_\beta=7.820(70)$~MeV~\cite{87Gr.A}, a $2.3~\sigma$ difference from our value of $7.984(7)$~MeV.  The AME03 mass excess value of $-72.330(80)$~MeV disagrees with the CPT value of $-71.9798(60)$~MeV, a larger \hbox{$4.4$-$\sigma$}, \hbox{350-keV} difference.  Of this difference, 186~keV is accounted for by our disagreement on the mass of $^{138}$Xe discussed below.

\textbf{$^{139}$I:} A single previous measurement was available for the AME03, a $\beta$-endpoint measurement that gives $Q_\beta=6.806(23)$~MeV~\cite{92Gr06}, a \hbox{379-keV}, \hbox{$9.8$-$\sigma$} difference from the CPT result of $7.185(31)$~MeV.  Curiously, both the $^{139}$Xe and $^{139}$I $Q_\beta$ measurements that the AME03 relies upon are in this same paper but have wildly different agreement with our results.  The recent FRS-ESR mass excess result of $-68.527(121)$~MeV~\cite{GSI_08} agrees with the CPT result of $-68.460(29)$~MeV.

\textbf{$^{137}$Xe:} By virtue of the adjacency of $^{137}$Xe to the stable $^{136}$Xe, the mass of $^{137}$Xe can be derived quite precisely from ($n,\gamma$) measurements.  The AME03 cites such a measurement in a draft IAEA Technical Document.  That document has since been published, with $S_n=4025.53(8)$~keV~\cite{Tecdoc}, consistent with the CPT result of $4015(11)$~keV.

Another Penning trap, ISOLTRAP, recently measured all xenon isotopes from $A=136$ to $146$ with similar precision as these measurements~\cite{ISOL-Xe09}.  The reported ISOLTRAP mass excess for $^{137}$Xe of $-82.3822(18)$~MeV is in agreement with the CPT value of $-82.373(11)$~MeV.

\textbf{$^{138}$Xe:} Two $\beta$-endpoint measurements are used for the mass determination by the AME03.  Those results are $Q_\beta = 2.720(50)$~MeV~\cite{72Mo33} and $2.830(80)$~MeV~\cite{78Wo15}.  These are combined in the AME03 to give the adopted value of $2.740(40)$~MeV.  As the mass of $^{138}$Cs was not measured in our experiment, a calculation of $Q_\beta$ entirely from CPT data is not possible.  It was previously measured by ISOLTRAP, however, with an adjusted mass excess of $-82.887(13)$~MeV~\cite{99Am05} (see $^{141}$Cs, below, for adjustment).  Combining these two Penning-trap measurements gives $Q_\beta = 2.923(14)$~MeV: a \hbox{183-keV}, \hbox{$4.3$-$\sigma$} disagreement.

ISOLTRAP's recent measurement of $^{138}$Xe with mass excess $-79.9751(33)$~MeV~\cite{ISOL-Xe09} differs from the CPT value of $-79.9641(55)$~MeV by $1.7~\sigma$.

\textbf{$^{139}$Xe:} Two experiments have measured $Q_\beta$ for this nuclide and are used by the AME03: $5.020(60)$~\cite{78Wo15} and $5.062(22)$~MeV~\cite{92Gr06}, averaged in the AME03 to $5.057(21)$~MeV.  Using the adjusted mass excess value of $^{139}$Cs from ISOLTRAP of $-80.704$~MeV~\cite{99Am05} (see $^{141}$Cs, below, for adjustment), our $Q_\beta=5.059(17)$~MeV is consistent.  The recent ISOLTRAP measurement of the $^{139}$Xe mass excess is $-75.6446(21)$~MeV~\cite{ISOL-Xe09}, which is also consistent with the CPT value of $-75.645(11)$~MeV.

\textbf{$^{140}$Xe:} There is only one existing $Q_\beta$ measurement for $^{140}$Xe, of $4.060(60)$~MeV~\cite{78Wo15}.  Using the adjusted ISOLTRAP $^{140}$Cs mass excess result of $-77.046$~MeV~\cite{99Am05} (see $^{141}$Cs, below, for adjustment), the CPT $Q_\beta=4.070(13)$~MeV agrees. The CPT mass excess of $-72.976(10)$~MeV is consistent with both the ISOLTRAP measurement of $-72.9865(23)$~MeV~\cite{ISOL-Xe09} and FRS-ESR measurement of $-72.870(121)$~MeV~\cite{GSI_08}.

\textbf{$^{141}$Xe:}  This xenon isotope also has a single $Q_\beta$ measurement, with $Q_\beta=6.150(90)$~MeV~\cite{78Wo15}.  The CPT result of $Q_\beta=6.266(21)$~MeV differs by $1.3~\sigma$.  The CPT mass excess result of $-68.1995(94)$~MeV is consistent with ISOLTRAP's result of $-68.1973(29)$~MeV~\cite{ISOL-Xe09}.  The recent FRS-ESR measurement disagrees by $2.5~\sigma$, with mass excess $-68.521(127)$~MeV~\cite{GSI_08}.

\textbf{$^{141}$Cs:} Neutron-rich cesium isotopes up to $^{142}$Cs~\cite{99Am05} as well as $^{145,147}$Cs~\cite{ISOL-Cs08} have already been measured quite precisely in ISOLTRAP, and to $^{148}$Cs with the Orsay double-focusing mass spectrometer~\cite{86Au02}, also at ISOLDE.  Therefore cesium was not investigated extensively in this experiment, with only $^{141,142}$Cs measured as a consistency check.

For $^{141}$Cs, the AME03 quotes three measurements.  Most heavily weighted is the ISOLTRAP measurement of the mass~\cite{99Am05}, which used $^{133}$Cs as the calibrant.  The accepted mass of that calibrant had increased between the original paper and the AME03 by $4.7$~keV$/c^2$.  Taking this into account the ISOLTRAP mass excess is $-74.475(15)$~MeV, consistent with our value of $-74.466(19)$~MeV.

The other two cited measurements are a $\beta$-endpoint measurement of $Q_\beta = 5.242(15)$~MeV~\cite{92Pr04}, and a beta-delayed neutron measurement~\cite{84Kr.B}.  We can compare with the $Q_\beta$ measurement by using an earlier CPT measurement of $^{141}$Ba~\cite{Savard_IJMS} to calculate $Q_\beta = 5.274(20)$~MeV, a difference of $1.3~\sigma$.   We can compare to the AME03 input $Q_{\beta n}$ value of $735(30)$~keV~\cite{84Kr.B} using the similarly adjusted ISOLTRAP mass for $^{140}$Ba~\cite{99Am05} which yields $752(25)$~keV, in agreement.

\textbf{$^{142}$Cs:} For this nuclide the AME03 considers the ISOLTRAP mass measurement~\cite{99Am05} and a $Q_\beta$ measurement~\cite{92Pr04} with an additional contribution from the Orsay mass spectrometer~\cite{86Au02}, with which we cannot make a direct comparison.  The adjusted (see $^{141}$Cs, above) ISOLTRAP mass excess of $-70.521(15)$~MeV is in agreement with our result of $-70.511(10)$~MeV.  Again utilizing previous CPT mass measurements~\cite{Savard_IJMS}, we can calculate $Q_\beta$ to be $7.340(15)$~MeV, which is 1.2~$\sigma$ from the value from \cite{92Pr04} of $7.315(15)$~MeV.

\textbf{$^{153}$Pr:}  This mass has previously been measured by the CPT~\cite{Savard_IJMS} at $152.933\,8895(153)$~u. This is in agreement with the value found in the present experiment of $152.933\,934(39)$~u.  The combined CPT result for this nuclide is therefore $152.933\,895(15)$~u.  The sole input for the AME03 is a $\beta$-endpoint measurement of $5.720(100)$~MeV~\cite{02Sh.B}, which agrees with our result of $5.756(15)$~MeV.

\textbf{$^{155}$Pr:}  There are no previous mass measurements of this nuclide, direct or indirect.  The AME03 extrapolates a mass excess of $-55.780(300)$~MeV~\cite{AME03}, which is $1.2~\sigma$ from our result of $-55.416(30)$~MeV.

\textbf{$^{153}$Nd:}  There is only one previous mass measurement of this neodymium isotope, a $\beta$-endpoint measurement of $Q_\beta=3.336(25)$~MeV~\cite{93Gr17}, which agrees with our result of $3.306(23)$~MeV.

\textbf{$^{155}$Nd:}  There is also only a single previous measurement of this isotope, a $\beta$-endpoint measurement~\cite{93Gr17} that was discarded by the AME03 due to severe disagreement with systematic trends.  That measurement was part of the same experiment as the accurate $^{153}$Nd measurement, above.  That result of $Q_\beta=4.222(150)$~MeV is inconsistent with our result of $4.656(18)$~MeV, which is closer to the AME03 interpolated value of $4.500(150)$~MeV~\cite{AME03}.

\textbf{$^{157}$Nd:}  The mass of this neodymium isotope has never before been measured by any means, directly or indirectly.  The AME03 extrapolates a mass excess of $-56.790(200)$~MeV~\cite{AME03}, $1.6~\sigma$ lighter than this work's $-56.464(43)$~MeV.

\textbf{$^{153}$Pm:}  Three measurements are listed in the AME03.  One is a $\beta$-endpoint $Q$-value experiment with a result of $Q_\beta~=~1.863(15)$~MeV~\cite{93Gr17}.  No measurement of $^{153}$Sm was made by the CPT, but using the well-established AME03 value for that mass the $Q_\beta$-value is found to be $1.928(23)$~MeV, a \hbox{2.4-$\sigma$} difference. A $^{154}$Sm$(d,^3$He$)^{153}$Pm $Q$-value measurement of $-3.623(25)$~MeV~\cite{76Sh.B} is also used.  Using the well-established mass of $^{154}$Sm~\cite{AME03} and our measurement, we find $Q_{(d,^3\textrm{He})}=-3.619(23)$~MeV, which is consistent.  A different proton-transfer reaction $Q$-value, $^{154}$Sm$(t,\alpha)^{153}$Pm~\cite{78Bu18}, was also measured at $10.748(20)$~MeV, which differs from our value of $10.701(25)$~MeV by $1.5~\sigma$.

\textbf{$^{155}$Pm:}  The only previous measurement is a $\beta$-endpoint measurement of $3.224(30)$~MeV~\cite{93Gr17}, and the CPT result of $3.254(24)$~MeV agrees.

\textbf{$^{156}$Pm:}  Two measurements are listed in the AME03.  Both are $\beta$-endpoint experiments~\cite{02Sh.B,90He11}, but the CPT has not measured $^{156}$Sm so a direct comparison of $Q_\beta$ is not possible.  However, the mass excess determination based on these measurements given by the AME03 of $-64.220(30)$~MeV differs from the CPT value of $-64.1655(59)$~MeV by $1.8~\sigma$.

\textbf{$^{157}$Pm:}  The AME03 cites a single $\beta$-endpoint measurement for this nuclide, with $Q_\beta=4.360(100)$~MeV~\cite{02Sh.B}.  The CPT measured both parent and daughter, finding $Q_\beta=4.380(15)$~MeV, which agrees.

\textbf{$^{158}$Pm:}  This has also been measured only by $\beta$-endpoint, with a sole experiment in the AME03 giving $Q_\beta=6.120(100)$~MeV~\cite{02Sh.A}.  That experiment has since been reanalyzed by the same group in Hayashi \emph{et al.}~\cite{Hayashi} resulting in $Q_\beta=6.085(80)$~MeV.  Both values agree with the CPT value of $6.160(26)$~MeV.

\textbf{$^{159}$Pm:}  This nuclide had not been measured by any means as of AME03 publication, but has since been subject of $\beta$-endpoint measurements~\cite{Shibata03} subsequently reanalyzed by Hayashi \emph{et al.} \cite{Hayashi} resulting in $Q_\beta~=~5.460(140)$~MeV.  This differs from the CPT value of $5.653(21)$~MeV by $1.4~\sigma$.  The AME03 systematic mass extrapolation gives a mass excess of $-56.850(200)$~MeV, which differs from the CPT value of $-56.557(17)$~MeV by $1.5~\sigma$.

\textbf{$^{155}$Sm:}  High-precision measurements of the neutron-separation energy of $^{155}$Sm have been made in two experiments via the $(n,\gamma)$ reaction, and these are used exclusively by the AME03 to determine its mass.  These measurements were adjusted in the AME03 to $S_n=5.8068(6)$~MeV~\cite{82Ba15} and $5.8070(3)$~MeV~\cite{82Sc03}.  Using the well-established AME03 value for the $^{154}$Sm mass, the CPT result is $S_n=5.806(22)$~MeV, which is consistent.

\textbf{$^{157}$Sm:}  A single $\beta$-endpoint measurement is used in the AME03 determination of this nuclide, with $Q_\beta=2.734(50)$~MeV~\cite{93Gr17}.  The CPT has not measured the daughter of this decay, so a direct comparison is impossible; however the CPT mass excess value of $-66.6795(74)$~MeV is consistent with the AME03 value of $-66.730(50)$~MeV.

\textbf{$^{158}$Sm:}  A single $\beta$-endpoint experiment forms the basis for the AME03 determination of this nuclide's mass, with $Q_\beta=1.999(15)$~MeV~\cite{93Gr17}, consistent with the CPT value of $2.010(27)$~MeV.

\textbf{$^{159}$Sm:}  The AME03 cites a single $\beta$-endpoint measurement for this nuclide of $Q_\beta=3.840(100)$~MeV~\cite{02Sh.A}, which has since been reanalyzed by the same group in Hayashi \emph{et al.} \cite{Hayashi} resulting in $Q_\beta=3.805(65)$~MeV.  Both values agree with the CPT value of $3.834(14)$~MeV.

\textbf{$^{160}$Sm:}  No previous mass measurement of any kind exists for $^{160}$Sm.  The AME03 extrapolated a mass excess of $-60.420(200)$~MeV, which is consistent with the CPT value of $-60.237(10)$~MeV.

\textbf{$^{161}$Sm:}  No measurement of this nuclide existed as of AME03 publication.  It has since been subject of the $\beta$-endpoint measurements~\cite{Shibata03} subsequently reanalyzed by the same group in Hayashi \emph{et al.} \cite{Hayashi} which found $Q_\beta=5.065(130)$~MeV, which agrees with the CPT result of $5.119(23)$~MeV.  The CPT mass measurement differs by 310~keV$/c^2$ with the AME03's extrapolation, which has an uncertainty of $300$~keV$/c^2$.

\textbf{$^{158}$Eu:}  This nuclide has two separate $\beta$-endpoint measurements, listed in the AME03 as $Q_\beta=3.550(120)$~\cite{65Sc19} and $3.440(100)$~MeV~\cite{66Da06}, combined in the AME03 to $3.490(80)$~MeV.  All of these are consistent with the CPT value of $3.435(25)$~MeV, using the well-established AME03 mass value for $^{158}$Gd which was not measured by the CPT.

\textbf{$^{159}$Eu:}  This nuclide was the subject of a transfer-reaction experiment, $^{160}$Gd$(t,\alpha)$, which gave a $Q$~value for that reaction of $10.636(8)$~MeV~\cite{79Bu05}.  Using the well-known mass of the nearly stable $^{160}$Gd from the AME03, the CPT measurement gives $Q_{(t,\alpha)}=10.622(9)$~MeV, a difference of $1.2~\sigma$.

\textbf{$^{160}$Eu:}  No previous mass measurements for this nuclide were used in the AME03, it having rejected two beta-endpoint measurements of $Q_\beta=3.900(300)$~\cite{73Da05} and $4.200(200)$~MeV~\cite{73Mo18}.  It has since been the subject of the $\beta$-endpoint measurements by Hayashi \emph{et al.} which found $Q_\beta=4.705(60)$~MeV~\cite{Hayashi}.  Using the well-known mass of the nearly stable $^{160}$Gd from the AME03, the CPT measurement disagrees significantly, with $Q_\beta=4.467(17)$~MeV, which is $238$~keV or $3.8~\sigma$ from the Hayashi \emph{et al.} result.  The CPT mass value is consistent with the AME03 systematic extrapolation.

\textbf{$^{161}$Eu:}   No previous mass measurement existed for this nuclide as of AME03 publication, but it has since been the subject of the $\beta$-endpoint measurements by Hayashi \emph{et al.} which found $Q_\beta=3.705(60)$~MeV~\cite{Hayashi}.  The CPT mass value is consistent with the AME03 systematic extrapolation.

\textbf{$^{163}$Gd:}  No previous mass measurement existed for this nuclide as of AME03 publication, but it has since been the subject of the $\beta$-endpoint measurements by Hayashi \emph{et al.} which found $Q_\beta=3.170(70)$~MeV~\cite{Hayashi}.  The CPT mass value is consistent with the AME03 systematic extrapolation.

\subsection{Mass Model Comparison}
\begin{figure}[h!]
\resizebox{0.49\textwidth}{!}{
\includegraphics{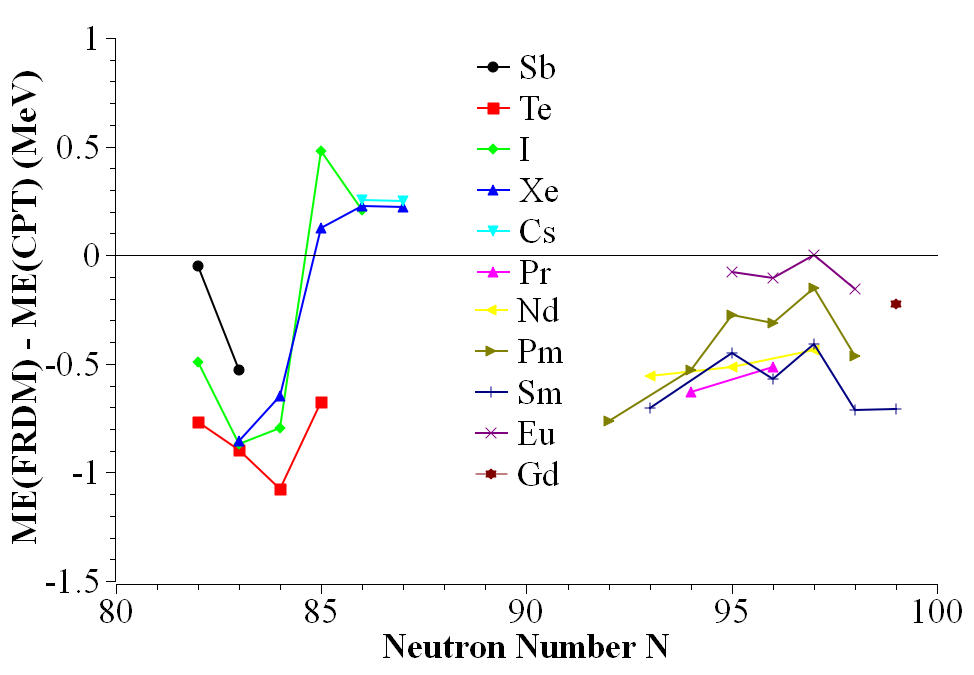}}
\caption{(Color online) Comparison between CPT measurements and FRDM~\cite{FRDM95} mass model calculations.}
\label{fig:FRDM-ME}
\end{figure}

\begin{figure}[h!]
\resizebox{0.49\textwidth}{!}{
\includegraphics{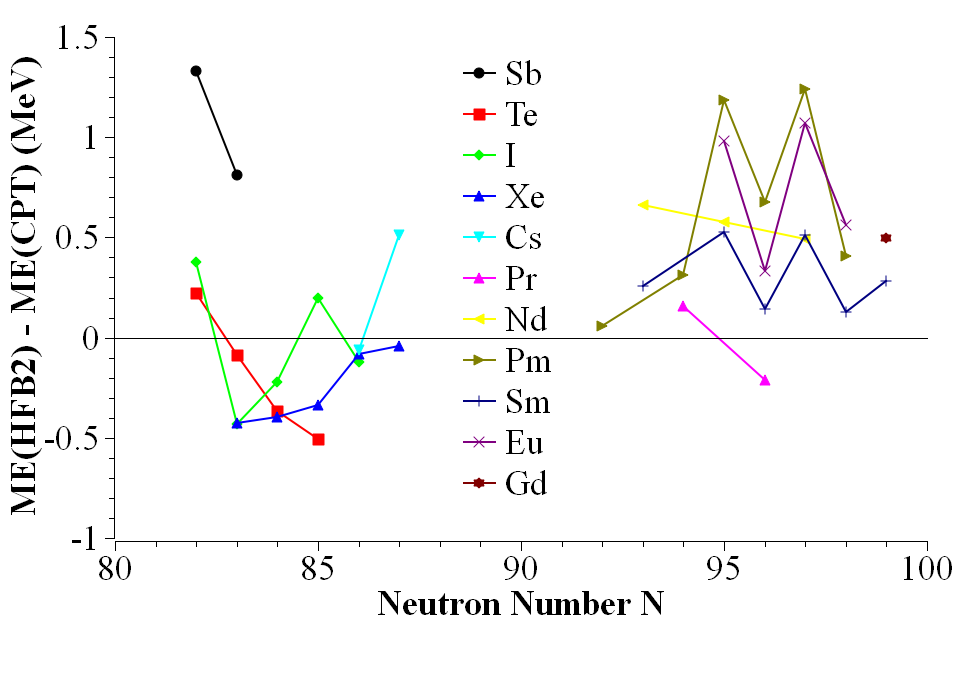}}
\caption{(Color online) Comparison between CPT measurements and HFB2~\cite{HFB2} mass model calculations.}
\label{fig:HFB2-ME}
\end{figure}

\begin{figure}[h!]
\resizebox{0.49\textwidth}{!}{
\includegraphics{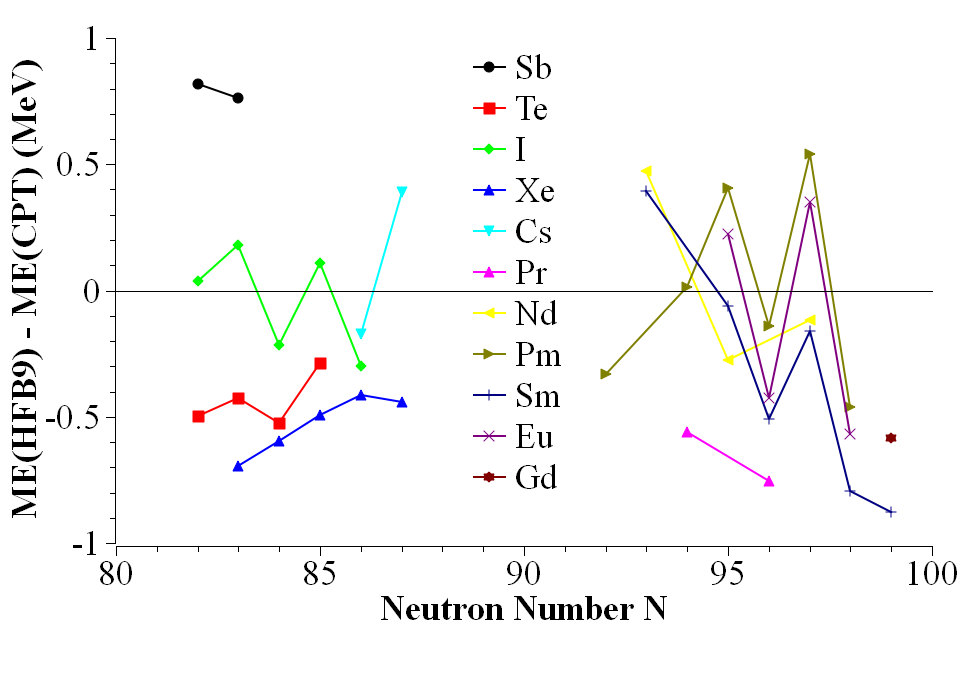}}
\caption{(Color online) Comparison between CPT measurements and HFB9~\cite{HFB9} mass model calculations.}
\label{fig:HFB9-ME}
\end{figure}

\begin{figure}[h!]
\resizebox{0.49\textwidth}{!}{
\includegraphics{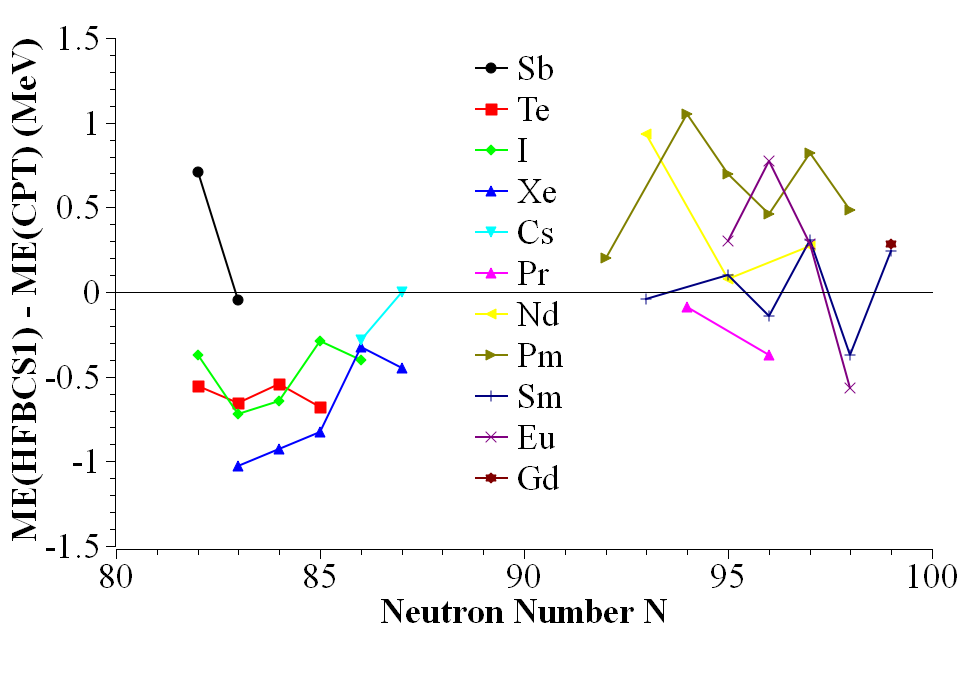}}
\caption{(Color online) Comparison between CPT measurements and HFBCS1~\cite{HFBCS1} mass model calculations.}
\label{fig:HFBCS1-ME}
\end{figure}

\begin{figure}[h!]
\resizebox{0.49\textwidth}{!}{
\includegraphics{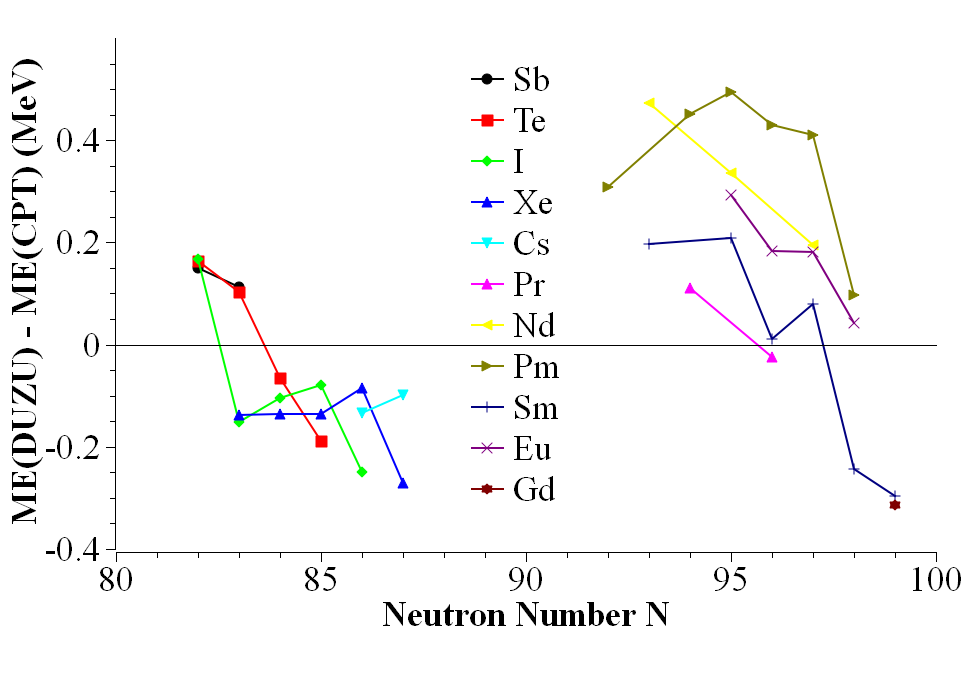}}
\caption{(Color online) Comparison between CPT measurements and DUZU~\cite{DUZU28} mass model calculations.}
\label{fig:DUZU-ME}
\end{figure}

\begin{figure}[h!]
\resizebox{0.49\textwidth}{!}{
\includegraphics{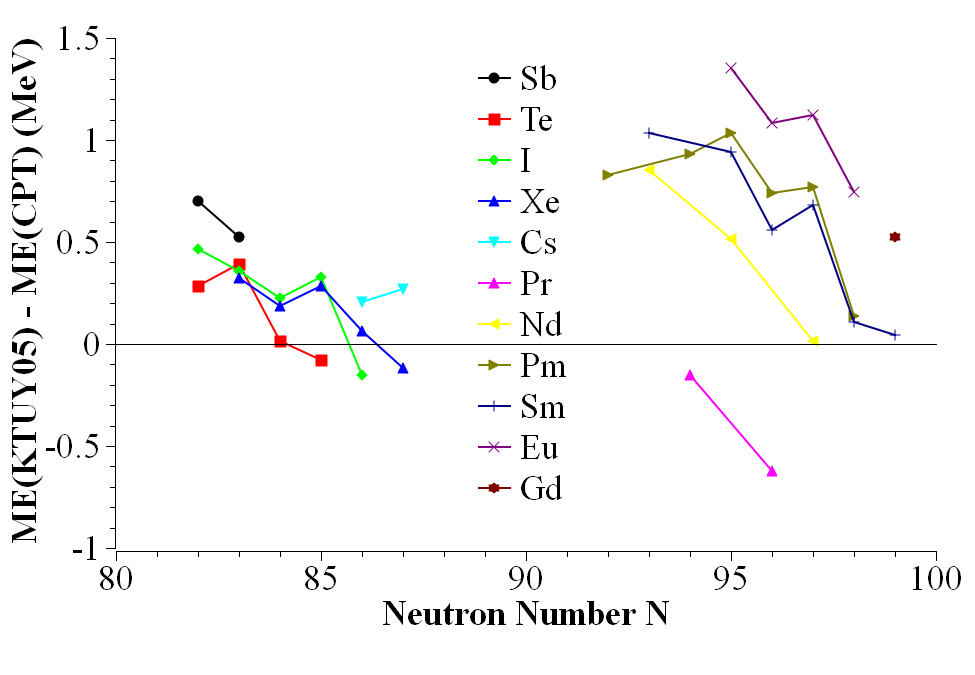}}
\caption{(Color online) Comparison between CPT measurements and KTUY05~\cite{KTUY05} mass model calculations.}
\label{fig:KTUY05-ME}
\end{figure}

\begin{figure}[h!]
\resizebox{0.49\textwidth}{!}{
\includegraphics{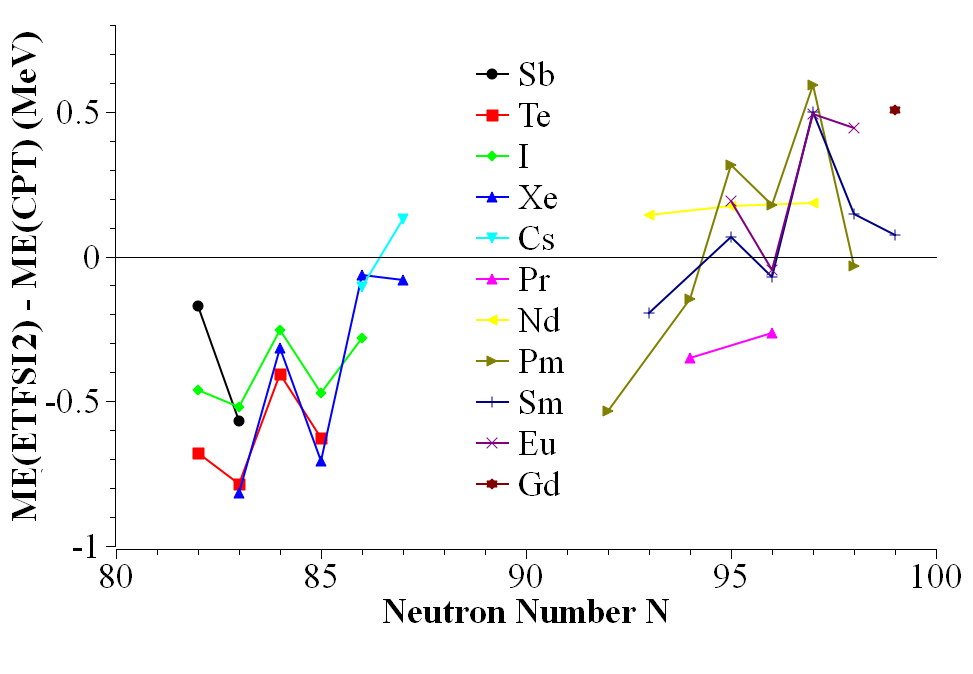}}
\caption{(Color online) Comparison between CPT measurements and ETFSI2~\cite{ETFSI2} mass model calculations.}
\label{fig:ETFSI12-ME}
\end{figure}

\begin{center}
\begin{table}[t!]
\begin{tabular}{l@{\hspace{0.5cm}} c@{\hspace{0.5cm}} r@{.}l}
\hline\noalign{\smallskip}
Model & $\sigma$ (MeV) & \multicolumn{2}{c}{$\bar{\epsilon}$ (MeV)} \\
\noalign{\smallskip}\hline\noalign{\smallskip}
AME03~\cite{AME03} & 0.171 & -0&105 \\
FRDM~\cite{FRDM95} & 0.538 & -0&379 \\
HFB2~\cite{HFB2} & 0.555 & 0&281 \\
HFB9~\cite{HFB9} & 0.467 & -0&175 \\
HFBCS1~\cite{HFBCS1} & 0.546 & -0&025 \\
DUZU~\cite{DUZU28} & 0.234 & 0&062 \\
KTUY05~\cite{KTUY05} & 0.611 & 0&438 \\
ETFSI2~\cite{ETFSI2} & 0.396 & -0&120 \\
\noalign{\smallskip}\hline
\end{tabular}
\caption{The RMS mass-excess difference ($\sigma$) and mean mass-excess difference ($\bar{\epsilon}$) of various mass models and the AME03 from the CPT for the measured nuclides presented here.}
\label{tbl:models}
\end{table}
\end{center}
Because most of the $r$-process path is outside the region of known masses, $r$-process simulations are forced to use mass models for neutron-separation energy values.  Available mass models often disagree significantly for unmeasured masses, so it can be enlightening to compare first-time measurements to mass models which were created before these masses were available.  Table~\ref{tbl:models} summarizes the overall precision and accuracy of, and  Figs.~\ref{fig:FRDM-ME}\textendash\ref{fig:ETFSI12-ME} compare the new CPT measurements to, various popular mass models.

The FRDM is a commonly used model for $r$-process simulations, and a comparison with the new CPT results is shown in Fig.~\ref{fig:FRDM-ME}.  The $N=82$ shell closure is followed by an oscillation about the true masses at higher $N$, though these oscillations appear to be damped away by $N=95$.  This region is important for the $r$ process, because its path is expected to exit the $N=82$ shell after the possible $^{130}$Cd waiting point~\cite{Cowan-rev}. If the overshooting in binding energy past $N=82$ continues for $Z$ lower than measured here, then the location of the $r$-process path may be closer to stability for $Z=49$ and 50 than the FRDM would indicate.  Clearly more mass measurements around $Z=49$ and 50 are vital for an accurate path determination past $N=82$.
\begin{figure*}[t!]
\begin{center}
\resizebox{0.9\textwidth}{!}{
\includegraphics{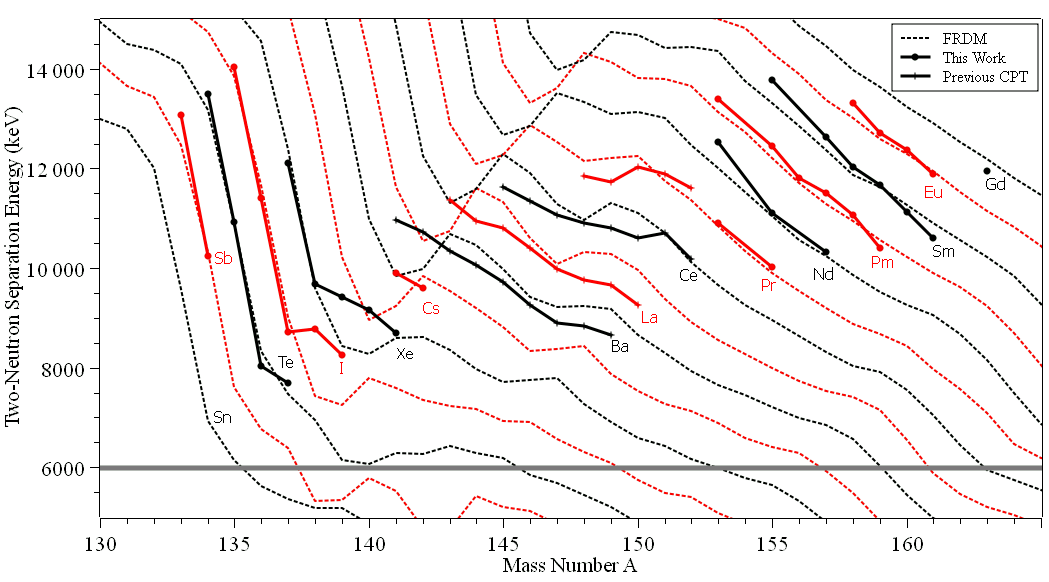}}
\caption{(Color online) Comparison between two-neutron separation energies for CPT measurements and the FRDM mass model calculations which have frequently been used for astrophysics simulations.  Colors alternate with Z for clarity.  The first two points for each element include one AME03 mass value each to determine $S_{2n}$, while the rest use only CPT data.  The crosses represent previously published CPT measurements~\cite{Savard_IJMS}.  The $r$-process path is thought to lie near $S_{2n}\approx 6000$~keV, as indicated by the grey line.}
\label{fig:FRDM_s2n}
\end{center}
\end{figure*}

For astrophysics modeling it is not the absolute errors in mass models that are relevant but systematically increasing errors with $N$, because this is what affects $S_n$ and thus the process paths.  For example, Table~\ref{tbl:models} would suggest that the DUZU model is one of the most accurate in this region, but Fig.~\ref{fig:DUZU-ME} shows that for most elements the deviations from our results have negative slope at $50$ to $100$ keV per neutron.  Given the $S_n$ slope of DUZU here, the difference in $S_n$ corresponds to a shift in the $r$-process path of about one neutron closer to stability were the $S_n$ difference to continue changing with the same slope.  KTUY05 (Fig.~\ref{fig:KTUY05-ME}) shows the same trend but with even larger magnitude. ETFSI2 (Fig.~\ref{fig:ETFSI12-ME}) on the other hand has a slope of similar magnitude in the opposite direction.  Any such trend in HFB2 or HFB9 (Fig.~\ref{fig:HFB2-ME} and ~\ref{fig:HFB9-ME}) is harder to identify given the apparent difficulty with pairing effects at large $N$, with many changes of over $500$~keV in mass deviation from isotope to isotope.  The HFBCS1 model (Fig.~\ref{fig:HFBCS1-ME}) had the lowest mean mass difference from the CPT among these models, but had among the largest scatters in difference, spanning over 2 MeV.

Fig.~\ref{fig:FRDM_s2n} compares neutron separation energies for the CPT and FRDM and suggests a more complicated story for this model. Of course these new measurements are still very far from the predicted $r$-process path, so reliable predictions of the path location are not yet possible.  Given the varied problems with mass models, forming any expectation that one would be more accurate than the others at yet higher neutron excess would be a gamble.  It is clear that mass measurements on or close to the $r$-process path are sorely needed, especially at the $N=82$ waiting points.

\section{\label{sec:Summary}Summary and Outlook}

Mass measurements of 40 nuclides were made, most of which improved the precision and accuracy over literature values.  Results are largely consistent with previous results from Penning traps, reaction energetics, and storage ring measurements, but are frequently much lower in mass than results from $\beta$-endpoint measurements.  Comparison with mass models shows only sporadic agreement, and suggests unsuitability of these models for precise astrophysical $r$-process simulations.  Reliable mass measurements yet closer to and on the $r$-process path are still needed.

The CPT has been moved to the new CAlifornium Rare Isotope Breeder Upgrade (CARIBU) facility~\cite{CARIBU} now operating at the ATLAS accelerator, and has begun a successor campaign to these measurements.  The extent of the CPT's previous measurements was limited by the purity and rate of the available beam, and CARIBU offers several improvements over the system used here to overcome these limitations.  The $10^3$-$10^4$ times more intense fission source and larger gas catcher will allow access to nuclides 3-5 neutrons farther from stability.  A new compact higher-resolution in-flight isobar separator~\cite{CARIBU_IS} promises purification with a resolving power of \hbox{20\,000}---four times that of the previous system---while avoiding the saturation of ion traps that would have occurred otherwise with such an intense beam.  The in-flight separation will also defer the limitation in nuclide lifetime that would be imposed by in-trap purification.   Many of the newly available nuclides lie directly on potential $r$-process paths near $Z=50$.  The CPT will conduct a survey of all accessible $r$-process nuclides in this vicinity, as well as any other neutron-rich species made available with the goal of finding isomers, regions of deformation, and shell quenching.  Over 200 additional masses are expected to be measured by the CPT.

\section{Acknowledgments}

This work performed under the auspices of NSERC,
Canada, application number 216974, and the U.S. DOE,
Office of Nuclear Physics, under Contract Nos. DE-AC02-06CH11357, DE-FG02-91ER-40609, DE-FG02-98ER41086,
and DE-AC52-07NA27344.


\begin{thebibliography}{99}

\bibitem{Savard_IJMS} G. Savard \emph{et al.}, Int. J. Mass Spectrom. {\bf 251}, 252 (2006).

\bibitem{SHIPTRAP-146} C. Rauth \emph{et al.}, Eur. Phys. J. Special Topics {\bf 150}, 329 (2007).

\bibitem{99Am05} F. Ames \emph{et al.}, Nucl. Phys. {\bf A651}, 3 (1999).

\bibitem{ISOL-Cs08} C. Weber, G. Audi, D. Beck, K. Blaum, G. Bollen, F. Herfurth, A. Kellerbauer, H.-J. Kluge, D. Lunney, and S. Schwarz, Nucl. Phys. {\bf A803}, 1 (2008).

\bibitem{Jyfl_Sr} U. Hager \emph{et al.}, Phys. Rev. Lett. {\bf 96}, 042504 (2006).

\bibitem{Jyfl_Tc} U. Hager \emph{et al.}, Phys. Rev. C {\bf 75}, 064302 (2007).

\bibitem{ISOL-Xe09} D. Neidherr \emph{et al.}, Phys. Rev. C {\bf 80}, 044323 (2009).

\bibitem{TITAN-12Be} S. Ettenauer \emph{et al.}, Phys. Rev. C {\bf 81}, 024314 (2010).

\bibitem{Bollen-review} G. Bollen and S. Schwarz, J. Phys. B {\bf 36}, 941 (2003).

\bibitem{Blaum-review} K. Blaum, Yu.N. Novikov, and G. Werth, Contemp. Phys. {\bf 51}, 149 (2010).

\bibitem{B2FH} E. M. Burbidge, G. R. Burbidge, W. A. Fowler, and F. Hoyle, Rev. Mod. Phys. {\bf 29}, 547 (1957)

\bibitem{Cowan-rev} J. J. Cowan, F.-K. Thielemann, J. W. Truran, Phys. Rep. {\bf 208}, 267 (1991).

\bibitem{Arnould-rev} M. Arnould, S. Goriely, and K. Takahashi, Phys. Rep. {\bf 450}, 97 (2007).

\bibitem{Arcones_simulation} A. Arcones and G. Mart\'{i}nez-Pinedo, Phys. Rev. C {\bf 83}, 045809 (2011).

\bibitem{Farouqi_simulation} K. Farouqi, K.-L. Kratz, B. Pfeiffer, T. Rauscher, F.-K. Thielemann, and J. W. Truran, Astrophys. J. {\bf 712}, 1359 (2010).

\bibitem{Wanajo_simulation} S. Wanajo, S. Goriely, M. Samyn, and N. Itoh, Astrophys. J. {\bf 606}, 1057 (2004).

\bibitem{deformation} R. Fossion, C. De Coster, J. E. Garc\'{i}a-Ramos, T. Werner, and K. Heyde, Nucl. Phys. {\bf A697}, 703 (2002).

\bibitem{quench} I. Dillmann \emph{et al.}, Phys Rev. Lett. {\bf 91}, 162503 (2003).

\bibitem{Fallis_thesis} J. Fallis, Ph.D. thesis, University of Manitoba, 2009 [http://hdl.handle.net/1993/3210].

\bibitem{Savard_catcher} G. Savard \emph{et al.}, Nucl. Instrum. Methods B {\bf 204}, 582 (2003).

\bibitem{Wada} M. Wada \emph{et al.}, Nucl. Instrum. Methods B {\bf 204}, 570 (2003).

\bibitem{Herfurth} F. Herfurth \emph{et al.}, Nucl. Instrum. Methods A {\bf 469}, 254 (2001).

\bibitem{Savard_gas} G. Savard, St. Becker, G. Bollen, H.-J. Kluge, R. B. Moore, Th. Otto, L. Scweikhard, H. Stolzenberg, and U. Wiess, Phys. Lett. A {\bf 158}, 247 (1991).

\bibitem{Geonium} S. Brown and G. Gabrielse, Rev. Mod. Phys. {\bf 58}, 233 (1986).

\bibitem{Bollen_TOF} G. Bollen, R. B. Moore, G. Savard, and H. Stolzenberg, J. Appl. Phys. {\bf 68}, 4355 (1990).

\bibitem{Konig} M. K\"{o}nig, G. Bollen, H.-J. Kluge, T. Otto, and J. Szerypo, Int. J. Mass Spectrom. Ion Proc. {\bf 142}, 95 (1995).

\bibitem{Clark_thesis} J. A. Clark, Ph.D. thesis, University of Manitoba, 2005 [http://hdl.handle.net/1993/177].

\bibitem{FSU_Xe} M. Redshaw, E. Wingfield, J. McDaniel, and E. G. Myers, Phys. Rev. Lett. {\bf 98}, 053003 (2007).

\bibitem{AME03} A. H. Wapstra, G. Audi, and C. Thibault, Nucl. Phys. A {\bf 729}, 129 (2003);  G. Audi, A. H. Wapstra, and C. Thibault, \emph{ibid.} {\bf 729}, 337 (2003).

\bibitem{Benzyne_heat} P. G. Wenthold, J. A. Paulino, and R. R. Squires, J. Am. Chem. Soc. {\bf 113}, 7414 (1991).

\bibitem{Benzyne_ion} X. Zhang and P. Chen, J. Am. Chem. Soc. {\bf 114}, 3147 (1992).

\bibitem{Chem_ref} M. W. Chase, \emph{NIST-JANAF Thermochemical Tables} (American Chemical Society, Washington, 1998), Vol. 1, p. 551, Vol 2, p. 1261.

\bibitem{Scielzo_Te} N. D. Scielzo \emph{et al.}, Phys. Rev. C {\bf 80}, 025501 (2009).

\bibitem{Gabby} G. Gabrielse, Int. J. Mass Spectrom. {\bf 279}, 107 (2009).

\bibitem{Bollen_isomer} G. Bollen, H.-J. Kluge, M. K\"{o}nig, T. Otto, G. Savard, H. Stolzenberg, R. B. Moore, G. Rouleau, G. Audi, and the ISOLDE Collaboration, Phys. Rev. C {\bf 46}, R2140 (1992).

\bibitem{02Ko53} A. Korgul \emph{et al.}, Eur. Phys. J. A {\bf 15}, 181 (2002).

\bibitem{Shergur} J. Shergur, A. W\"{o}hr, W. B. Walters, K.-L. Kratz, O. Arndt, B. A. Brown, J. Cederkall, I. Dillmann, L. M. Fraile, P. Hoff, A. Joinet, U. K\"{o}ster, and B. Pfeiffer, Phys. Rev. C {\bf 71}, 064321 (2005).

\bibitem{80KeZQ} U. Keyser, H. Berg, F. M\"{u}nich, B. Pahlman, K. Hawerkamp, B. Pfeiffer, H. Schrader, and E. Monnand, in \emph{Atomic Masses and Fundamental Constants, 6}, edited by J. A. Nolen, Jr. and W. Benenson (Plenum Press, New York, 1980), p. 485.

\bibitem{Fallis_90} J. Fallis \emph{et al.}, Phys. Rev. C {\bf 84}, 045807 (2011).

\bibitem{GSI_08} B. Sun \emph{et al.}, Nucl. Phys. {\bf A812}, 1 (2008).

\bibitem{Tecdoc} R. B. Firestone, S. F. Mughabghab, and G. L. Moln\'{a}r, IAEA Tecdoc, STI/PUB/1263 (2007).

\bibitem{82Ba15} I. F. Barchuk, V. I. Golyshkin, and E. N. Gorban, Izv. Akad. Nauk SSSR, Ser. Fiz. {\bf 46}, 63 (1982).

\bibitem{82Sc03} K. Schreckenbach \emph{et al.}, Nucl. Phys. {\bf A376}, 149 (1982).

\bibitem{71Wi04} B. H. Wildenthal, E. Newman, and R. L. Auble, Phys. Rev. C {\bf 3}, 1199 (1971).

\bibitem{76Sh.B} E. Sugarbaker and W. S. Gray, Bull. Am. Phys. Soc. {\bf 21}, 984 (1976).

\bibitem{78Bu18} D. G. Burke, G. L{\o}vh{\o}iden, E. R. Flynn, and J. W. Sunier, Phys. Rev. C {\bf 18}, 693 (1978).

\bibitem{79Bu05} D. G. Burke, G. L{\o}vh{\o}iden, E. R. Flynn, and J. W. Sunier, Nucl. Phys. {\bf A318}, 77 (1979).

\bibitem{Hayashi} H. Hayashi \emph{et al.}, Eur. Phys. J. A {\bf 34}, 363 (2007).

\bibitem{02Sh.A} M. Shibata, T. Shindou, Y. Kojima, M. Asai, K. Tsukada, S. Ichikawa, H. Haba, Y. Nagame, and K. Kawade, Japan Atomic Energy Research Institute, Tandem VDG Annual Report, 2001 (2002) p. 26; JAERI-Review 2002-029 (2002).

\bibitem{Shibata03} M. Shibata, O. Suematsu, K. Kawade, M. Asai, S. Ichikawa, Y. Nagame, A. Osa, K. Tsukada, Y. Kojima, and A. Taniguchi, Japan Atomic Energy Research Institute, Tandem VDG Annual Report, 2002 (2003) p. 32; JAERI-Review 2003-028 (2003).

\bibitem{07Fo02} B. Fogelberg, K. A. Mezilev, V. I. Isakov, K. I. Erokhina, H. Mach, E. Ramstr\"{o}m, and H. Gausemel, Phys. Rev. C {\bf 75}, 054308 (2007).

\bibitem{Shibata_unc} M. Shibata, Y. Kojima, H. Uno, K. Kawade, A. Taniguchi, Y. Kawase, S. Ichikawa, F. Maekawa, and Y. Ikeda, Nucl. Instrum. Methods A {\bf 459}, 581 (2001).

\bibitem{99Fo01} B. Fogelberg, K. A. Mezilev, H. Mach, V. I. Isakov, and J. Slivova, Phys. Rev. Lett. {\bf 82}, 1823 (1999).

\bibitem{95Me16} K. A. Mezilev, Yu. N. Novikov, A. V. Popov, B. Fogelberg, and L. Spanier, Phys. Scr. {\bf T56}, 272 (1995).

\bibitem{87Gr.A} M. Graefenstedt, U. Keyser, F. M\"{u}nnich, and F. Schreiber, AIP Conf. Proc. {\bf 164}, 30 (1987).

\bibitem{85Sa15} M. Samri, G. J. Costa, G. Klotz, D. Magnac, R. Seltz, and J. P. Zirnheld, Z. Phys. A {\bf 321}, 255 (1985).

\bibitem{84Kr.B} K.-L. Kratz, A. Schr\"{o}eder, H. Ohm, H. Gabelmann, W. Ziegert, B. Steinm\"{u}ller, and B. Pfeiffer, in \emph{Proceedings of the 7th International Conference on Atomic Masses and Fundamental Constants AMCO-7}, edited by O. Klepper (Technishe Hochschule Darmstadt Lehrdruckerei, Damstadt, 1984), p. 127. 

\bibitem{77Sc21} F. Schussler, J. Blachot, E. Monnand, J. A. Pinston, and B. Pfeiffer, Z. Phys. A {\bf 283}, 43 (1977).

\bibitem{76Lu04} E. Lund and G. Rudstam, Nucl. Instrum. Methods {\bf 134}, 173 (1976).

\bibitem{59Jo37} N. R. Johnson and G. D. O'Kelley, Phys. Rev. {\bf 114}, 279 (1959).

\bibitem{92Gr06} M. Gro\ss, P. J\"{u}rgens, U. Keyser, S. Kluge, M. Mehrtens, S. M\"{u}ller, F. M\"{u}nnich, J. Wulff, and H. R. Faust, Nucl. Instrum. Methods A {\bf 311}, 512 (1992).

\bibitem{72Mo33} E. Monnand, R. Brissot, L. C. Carraz, J. Cran\c{c}on, C. Ristori, F. Schussler, and A. Moussa, Nucl. Phys. {\bf A195}, 192 (1972).

\bibitem{78Wo15} F. K. Wohn and W. L. Talbert, Jr., Phys. Rev. C {\bf 18}, 2328 (1978).

\bibitem{86Au02} G. Audi, A. Coc, M. Epherre-Rey-Campagnolle, G. Le Scornet, C. Thibault, and F. Touchard, Nucl. Phys. {\bf A449}, 491 (1986).

\bibitem{92Pr04} M. Przewloka, A. Przewloka, P. W\"{a}chter, and H. Wollnik, Z. Phys. A {\bf 342}, 27 (1992).

\bibitem{02Sh.B} M. Shibata, T. Shindou, K. Kawade, Y. Kojima, A. Taniguchi, Y. Kawase, and S. Ichikawa, in \emph{Exotic Nuclei and Atomic Masses, ENAM2001}, edited by J. \"{A}yst\"{o}, P. Dendooven, A. Jokinen, and M. Leino (Springer-Verlag, Berlin, Heidelberg, 2001), p. 479.

\bibitem{93Gr17} R. C. Greenwood and M. H. Putnam, Nucl. Instrum. Methods A {\bf 337}, 106 (1993).

\bibitem{90He11} M. Hellstr\"{o}m, B. Fogelberg, L. Spanier, and H. Mach, Phys. Rev. C {\bf 41}, 2325 (1990).

\bibitem{65Sc19} F. Schima and T. Katoh, Phys. Rev. {\bf 140}, B1496 (1965).

\bibitem{66Da06} W. R. Daniels and D. C. Hoffman, Phys. Rev. {\bf 147}, 845 (1966).

\bibitem{73Da05} J. M. D'Auria, R. D. Guy, and S. C. Gujrathi, Can. J. Phys. {\bf 51}, 686 (1973).

\bibitem{73Mo18} N. A. Morcos, W. D. James, D. E. Adams, and P. K. Kuroda, J. Inorg. Nucl. Chem. {\bf 35}, 3659 (1973).

\bibitem{FRDM95} P. M\"{o}ller, J. R. Nix, W. D. Myers, and W. J. Swiatecki, At. Data Nucl. Data Tables {\bf 59}, 185 (1995).

\bibitem{HFB2} S. Goriely, M. Samyn, P.-H. Heenen, J. M. Pearson, and F. Tondeur, Phys. Rev. C {\bf 66}, 024326 (2002).

\bibitem{HFB9} S. Goriely, M. Samyn, J. M. Pearson, and M. Onsi, Nucl. Phys. {\bf A750}, 425 (2005).

\bibitem{HFBCS1} F. Tondeur, S. Goriely, J. M. Pearson, and M. Onsi, Phys. Rev. C {\bf 62}, 024308 (2000).

\bibitem{DUZU28} J. Duflo and A. P. Zuker, Phys. Rev. C {\bf 52}, R23 (1995).

\bibitem{KTUY05} H. Koura, T. Tachibana, M. Uno, and M. Yamada, Prog. Theor. Phys. {\bf 113}, 305 (2005).

\bibitem{ETFSI2} S. Goriely, AIP Conf. Proc. {\bf 529}, 287 (2000).

\bibitem{CARIBU} G. Savard, S. Baker, C. Davids, A.F. Levand, E.F. Moore, R.C. Pardo, R. Vondrasek, B.J. Zabransky, and G. Zinkann, Nucl. Instrum. Methods B {\bf 266}, 4086 (2008).

\bibitem{CARIBU_IS} C. N. Davids, D. Peterson, Nucl. Instrum. Methods B {\bf 266}, 4449 (2008).


\end{thebibliography}

\end{document}